\documentclass[aps,prc,twocolumn,showpacs,preprintnumbers,
                          nofootinbib,float,longbibliography]{revtex4-1}
\usepackage{graphicx, fancybox}
\usepackage{amsmath,amssymb}
\usepackage[usenames, dvipsnames]{color}
\usepackage{url}
\usepackage{adjustbox}
\usepackage{slashed}
\usepackage{bm}
\usepackage{relsize}
\usepackage{color}
\usepackage{hyperref}
\usepackage{caption}
\usepackage{subcaption}

\usepackage{xcolor, soul}

\newcommand{\beq}{\begin{equation}}
\newcommand{\eeq}{\end{equation}}
\newcommand{\ret}{\mathrm{ret}}
\newcommand{\adv}{\mathrm{adv}}

\begin{document}

\title{The momentum broadening of energetic partons in an anisotropic plasma}
\begin{abstract}
The quark-gluon plasma produced in heavy-ion collisions is anisotropic throughout its evolution. This anisotropy changes the physics of jet-medium interaction, making it dependent on the momentum direction of the jet. In this paper we analyze transverse momentum broadening of a jet parton interacting with soft gluons in an anisotropic plasma. Our analysis equally applies to momentum broadening of quasiparticles in kinetic theory. We subtract contribution from instability modes in the deep infrared and discuss how our calculation should be complemented in that regime. The resulting anisotropic collision kernel for momentum broadening is qualitatively different from the equilibrium collision kernel and from the isotropic ansatz used in effective kinetic theory. Because of increased medium screening, there is substantially less transverse broadening at low and intermediate momenta. 
\end{abstract}

\author{Sigtryggur Hauksson, Sangyong Jeon, Charles Gale}
\affiliation{Department of Physics, McGill University, 3600 University Street, Montr\'eal, QC, Canada H3A 2T8 \\
{\rm Correspondence}: sigtryggur.hauksson@mail.mcgill.ca}

\maketitle

\section{Introduction}
\label{sec:Introduction}

Heavy-ion collisions at RHIC and the LHC produce extremely dense and energetic matter governed by the strong interaction: this matter is known as the quark-gluon plasma (QGP) \cite{Jacak:2012dx}. One of the primary experimental probes of the QGP are jets which are produced during the initial hard scattering of nuclei. As these jets traverse the droplet of QGP their structure is modified, leaving imprints of the medium on experimental observables. 

A great deal of theoretical effort has explored how nearly on-shell jet partons interact with a weakly coupled QGP medium. 
A jet parton receives repeated small momentum kicks from medium particles leading to diffusion in the parton's momentum transverse to its direction of motion. These kicks bring the jet parton slightly off shell, allowing it to radiate gluons through medium-induced branching, which changes the jet shower relative to vacuum.
In addition, jets lose energy when traversing the plasma as they interact with their radiation field, and low-energy jet partons are absorbed by the medium. 

Using thermal field theory, one can evaluate transverse momentum broadening in a weakly-coupled plasma. This is quantified by the collision kernel \(\mathcal{C}(\mathbf{q}_{\perp})\) which is the probability of  receiving a transverse kick of momentum \(\mathbf{q}_{\perp}\) from the medium.
Such microscopic calculations exist at leading order in perturbation theory \cite{Aurenche:2002pd} as well as at next-to leading order \cite{CaronHuot:2008ni}. The collision kernel for transverse momentum broadening has furthermore been evaluated non-perturbatively on the lattice using electrostatic QCD effective field theory, see e.g. \cite{Panero:2013pla} and \cite{Moore:2021jwe}.

Analytic calculations of the collision kernel have so far assumed a medium in local thermal equilibrium, or with local isotropy in momentum space \cite{Hauksson:2017udm}. However, we know  that there are sizable deviation from isotropy and thermal equilibrium at all stages of heavy-ion collisions \cite{Strickland:2017kux}.
This can have important effects on the phenomenology of jets:
an anisotropy in the local momentum distribution of quarks and gluons leads to directional dependence in jet evolution. Specifically, jet partons going through the same patch of QGP but travelling in different directions will have different rates of momentum broadening and of medium-induced splitting.

Having an anisotropic collision kernel for momentum broadening is not only important for jet physics, but also for formulating a kinetic theory of quarks and gluons \cite{Arnold:2002zm}. Such kinetic theories are used to describe early stages of heavy-ion collisions after the glasma phase and before the hydrodynamic phase \cite{Kurkela:2018vqr,Kurkela:2018wud}. One of the two main processes for quasiparticle interaction is gluon radiation by a quark or a gluon which is brought slightly off-shell by momentum broadening. Up until now, kinetic theory calculations have employed an isotropic ansatz for the collision kernel \cite{Kurkela:2015qoa, AbraaoYork:2014hbk} but consistency requires a non-equilibrium kernel. This could affect results of kinetic theory simulations. 
We finally note that
calculation of photon radiation through bremsstrahlung in an anisotropic medium relies on precisely the same non-equilibrium collision kernel \cite{Arnold:2001ba, Arnold:2001ms, Hauksson:2017udm}.

Understanding momentum broadening in an anisotropic plasma requires a detailed microscopic calculation.  In this paper we provide such a calculation in the hard thermal loop (HTL) regime where self-interaction of soft gluons can be ignored. Specifically, we consider a medium in which quark and gluon quasiparticles are distributed anisotropically in momentum space. The quasiparticles source soft gluons which propagate until they give the jet parton a transverse kick. This gives rise to a collision kernel \(\mathcal{C}(\mathbf{q}_{\perp})\) which depends not only on local properties of the medium but also on the direction of the jet parton.  Our calculation is at leading order in perturbation theory. We subtract instability modes \cite{Hauksson:2020wsm} corresponding to exponential growth in soft gluon density \cite{Mrowczynski:2016etf}, as they have become saturated during the kinetic and hydrodynamic stages we are interested in \cite{Berges:2013eia,Berges:2013fga,Berges:2014bba}.


We note that numerical simulations have measured momentum broadening of jet partons and heavy quarks in a variety of equilibrium and non-equilibrium situations. These include classical-statistical field theory \cite{Boguslavski:2020tqz}, an HTL setup with kinetic theory for quasiparicles and classical field theory for soft gluons \cite{Schenke:2008gg,Mrowczynski:2017kso,Dumitru:2007rp}, as well as the color-glass condensate, see \cite{Ipp:2020nfu, Ipp:2020mjc} and \cite{Carrington:2020sww}. Furthermore, momentum broadening has been measured on the lattice in a factorized approach assuming a single scattering off medium gluons \cite{Kumar:2020wvb}. The effect of instabilities on momentum broadening was furthermore assessed in \cite{Majumder:2009cf,Carrington:2016mhd}.
We also note that \cite{Sadofyev:2021ohn} evaluated the effect of temperature and density gradients on momentum broadening and jet splitting, assuming a medium composed of massive particles and working in an opacity expansion.
Our analytic approach complements these studies as it gives results which are independent of assumptions of numerical simulations. Furthermore, our results only depend on the instantaneous properties of the medium. We evaluate the full collision kernel which is needed for the rate of gluon radiation.

The paper is organized as follows: In Sec.  \ref{sec:broadening_energyloss} we discuss the physics of jet momentum broadening in detail and show how it differs microscopically from jet energy loss, in a non-equilibrium medium. In Sec. \ref{sec:Density of soft gluons} we calculate the density of soft gluons in an anisotropic medium. In Sec. \ref{sec:instabilities} we discuss our treatment of instabilities in an anisotropic plasma. Finally, results are presented in Sec. \ref{sec:Results}. Some additional details are provided in Appendices.

\section{Jet momentum broadening and energy loss}
\label{sec:broadening_energyloss}


As a jet parton passes through a plasma it loses energy and gains momentum transverse to its direction of motion. This happens through three different processes: hard two-to-two scattering with plasma constituents, medium-induced bremsstrahlung, and interaction with soft gluons \cite{Ghiglieri:2015zma}. In this work we focus on interaction with soft gluons which has been less studied in a non-equilibrium medium and which furthermore is the basis of medium-induced splitting.

In thermal equilibrium the physics of energy loss due to soft gluons is different from that of momentum broadening from soft gluons:  momentum broadening results from transverse kicks of gluons that exist in the medium and have been radiated by quasiparticles, while energy loss results from soft gluons that the jet parton itself radiates, and not gluons present in the medium.

It is important to establish that energy loss and momentum broadening differ in the same way in a non-equilibrium system, i.e. that momentum broadening is due soft gluons in the medium while energy loss is due to the soft gluon radiation field of the parton itself. We will show this using the real-time formalism \cite{Bellac:2011kqa,Ghiglieri:2020dpq}
which will furthermore establish our conventions. We focus on an energetic quark parton traversing the plasma but the argument for a gluon parton or a heavy quark is nearly identical.

\begin{figure}
    \centering
         \includegraphics[width=0.35\textwidth]{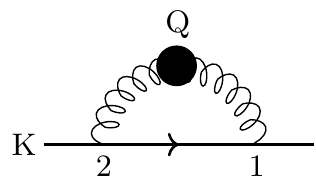}
         \caption{Self-energy diagram for a jet quark \(K\) interacting with a soft gluon \(Q\). The soft gluon propagator is resummed.}
         \label{Fig:feynm_fig}
\end{figure}

 A quark parton flying through the plasma with momentum \(K^{\mu} = (k^0,\mathbf{k})\) interacts with soft gluons. This leads to a decay rate \cite{Bellac:2011kqa}
 \beq
  \Gamma = \frac{1}{4k}\mathrm{Tr}\left[ \slashed{K}\Sigma_{21}\right] 
 \eeq
 where the self-energy component
 \beq
 \begin{split}
 \Sigma_{21}(K) &= g^2 C_F \int \frac{d^4 Q}{(2\pi)^4} \;D_{21}^{\mu\nu}(Q) \\
 &\times \gamma_{\mu} \left(\slashed{K} - \slashed{Q}\right) \gamma_{\nu} \left[1-f_{q}(\mathbf{k}-\mathbf{q})\right] 2\pi\delta((K-Q)^2).
 \end{split}
 \eeq
  is given by the diagram in Fig. \ref{Fig:feynm_fig}.
Using that the soft gluon momentum \(Q\) is much smaller than the jet parton momentum \(K\) we get a differential decay rate
\beq
\label{Eq:diff_Gamma}
\frac{d\Gamma}{d^4 Q} = \frac{g^2 C_F }{(2\pi)^4}\, D_{21}^{\mu\nu}(Q) v_{\mu} v_{\nu} \,\delta(v\cdot Q)
\eeq
where \(v_{\mu} = K^{\mu}/k\) is the normalized parton momentum. To generalize our results for a heavy quark, one simply substitutes \(v^{\mu}\) with \((k^0/k,\widehat{\mathbf{k}}) = (\sqrt{k^2+m^2}/k,\widehat{\mathbf{k}})\).
In either case, the essential ingredient is the resummed propagator for soft gluons, 
\beq
D_{21}^{\mu\nu}(Q) = \int d^4(x-y) \; e^{iQ\cdot(x-y)}\langle A^{\mu}(x) A^{\nu}(y)\rangle
\eeq which can be written more physically as\footnote{In a non-equilibrium plasma one uses a Wigner transform \(D_{21}^{\mu\nu}(Q,X) = \int d^4(x-y) \; e^{iQ\cdot(x-y)}\langle A^{\mu}(x) A^{\nu}(y)\rangle\) where \(X = (x+y)/2\) is the position in the plasma. In this section we assume that changes in \(X\) are slow and we omit writing dependence on \(X\) explicitly.} 
\beq
\label{Eq:D21_rewrite}
D^{\mu\nu}_{21}(Q) = \frac{1}{2}\left[D^{\mu\nu}_{\ret}(Q) - D^{\mu\nu}_{\adv}(Q) \right] + D^{\mu\nu}_{rr}(Q)
\eeq

The two terms in Eq. \eqref{Eq:D21_rewrite} have different interpretations. 
The \(rr\) propagator is 
\beq
D^{\mu\nu}_{rr}(x,y) = \frac{1}{4} \langle \{A^{\mu}(x), A^{\nu}(y)\}\rangle.
\eeq
It is non-vanishing even for classical fields, and describes the density of soft gluons in the medium. The contribution of the \(rr\) correlator to the decay rate in Eq. \eqref{Eq:diff_Gamma} thus corresponds to the jet parton interacting with soft gluons that are already present in the medium.

Meanwhile, the retarded correlator in Eq. \eqref{Eq:D21_rewrite} is given by
\beq
D^{\mu\nu}_{\ret}(x,y) = \theta(t_x-t_y) \langle \left[ A^{\mu}(x),A^{\nu}(y)\right] \rangle.
\eeq
It describes the causal propagation of soft gluons forward in time. Similarly, 
the advanced correlator is \(D_{\adv}(x,y) = -\theta(t_y-t_x) \langle \left[ A^{\mu}(x),A^{\nu}(y)\right] \rangle\).  The contribution of these correlators to the differential decay rate describes soft gluons radiated by the jet parton. These soft gluons propagate in the medium before giving the same parton a kick. In other words, in the regime of soft gluons the contribution of the retarded and advanced correlators does not represent gluons already present in the medium, but rather the interaction of the jet with its own radiation field.

The rate of energy loss of a jet parton due to soft gluons is
\beq
\begin{split}
\label{Eq:e_loss_deriv}
\widehat{e} &= \frac{d\langle p^0\rangle}{dt} = \int d^4 Q\; q^0 \frac{d\Gamma}{d^4 Q}
\end{split}
\eeq
Since \(D_{rr}^{\mu\nu}(x,y) = D_{rr}^{\nu\mu}(y,x)\), one gets that
\beq
D_{rr}^{\mu\nu}(Q)v_{\mu}v_{\nu} = D_{rr}^{\mu\nu}(-Q)v_{\mu}v_{\nu}. 
\eeq
Thus, the jet is equally likely to gain four-momentum \(Q\) from soft gluons in the medium as to lose four-momentum \(Q\). Therefore, 
\beq
\begin{split}
\label{Eq:e_loss}
\widehat{e} =  g^2 C_F \int \frac{d^4 Q}{(2\pi)^4}\; q^0   \frac{1}{2}\left(D^{\mu\nu}_{\ret} - D^{\mu\nu}_{\adv} \right) (Q) v_{\mu} v_{\nu} \,2\pi\delta(v\cdot Q).
\end{split}
\eeq
which is solely due to the jet's radiation field.

Eq. \eqref{Eq:e_loss} for energy loss has a simple physical interpretation given e.g. in \cite{Mrowczynski:1991da, Thoma:1990fm}.  The jet parton's radiation field is described by   classical field theory. The rate of energy loss is then  
\beq
\label{Eq:e_loss_class_der}
\widehat{e} = \mathrm{Re} \int d^3x\, \mathbf{J}_{\mathrm{ext}}(x) \cdot \mathbf{E}_{\mathrm{ind}}(x)
\eeq
where the current is the jet parton itself, \(\mathbf{J}_{\mathrm{ext}} = g \mathbf{v} \delta^{(3)}(\mathbf{x} - \mathbf{v} t)\), and \(\mathbf{E}_{\mathrm{ind}}\) is the electric field it induces. Linear response theory gives that 
\beq
\label{Eq:linear_response}
E^{i}_{\mathrm{ind}}(Q) = iq^0 D_{\ret}^{ij}(Q) J_{\mathrm{ext}}^{j}(Q).
\eeq
Combining these equations and taking care of color factors reproduces our formula for energy loss, Eq. \eqref{Eq:e_loss}. This shows that energy loss is indeed due to fields induced by the jet parton. \footnote{The expression given in \cite{Mrowczynski:1991da, Thoma:1990fm}
has \(D_{\ret} - D^0_{\ret}\) instead of \(D_{\ret}\) where \(D^0_{\ret}\) is the bare propagator. It is easy to see that the contribution of the bare propagator is phase space suppressed and can be omitted. Their expression furthermore assumes temporal axial gauge, \(A^0 = 0\). In a more general gauge, current conservation needs to be postulated explicitly for the classical argument to go through, see \cite{Carrington:2015xca}.
}

Eq. \eqref{Eq:e_loss} for energy loss has been evaluated in an anisotropic medium along with the contribution of hard medium particles \cite{Romatschke:2003vc,Romatschke:2004au}, see Appendix \ref{sec:Appendix_energyloss} for a discussion. Their formalism was based on the equilibrium calculation  in \cite{Braaten:1991jj,Braaten:1991we}. Jet energy loss has furthermore been evaluated in an evolving unstable plasma \cite{Carrington:2015xca}, see also \cite{Mrowczynski:2017kso}.

The physics of jet momentum broadening differs from that of energy loss.
In the HTL regime \(D_{rr}(Q) \sim 1/g^3\) because of high occupation density of soft gluons while \(D_{\ret} \sim 1/g^2\).  This can be seen immediately in thermal equilibrium  from Eq. \eqref{Eq:KMS_Drr}, and in an anisotropic system we will show this below. 
The rate of jet transverse momentum broadening is therefore
\beq
\label{Eq:qhat_def}
\begin{split}
\widehat{q} &= \frac{d\langle (\Delta p_{\perp})^2\rangle}{dt} = \int d^4 Q\; q^2_{\perp}\frac{d\Gamma}{d^4 Q} \\
&\approx  \int \frac{d^2 q_{\perp}}{(2\pi)^2}\; q^2_{\perp}\,\mathcal{C}(\mathbf{q}_{\perp})
\end{split}
\eeq
where the collision kernel for momentum broadening is 
\beq
\label{Eq:coll_kern}
\mathcal{C}(\mathbf{q}_{\perp}) = g^2 C_F \int \frac{dq^0 dq^z }{(2\pi)^2}\; D_{rr}^{\mu\nu}(Q) v_{\mu} v_{\nu} \,2\pi\delta(v\cdot Q)
\eeq
with \(z\) the direction of motion of the jet.
We see that jet momentum broadening is due to soft gluons already present in the medium and not due to the jet's radiation field.
The main task of this paper is to calculate the collision kernel \(\mathcal{C}(\mathbf{q}_{\perp})\) and the transport coefficient \(\widehat{q}\) in an anisotropic plasma.\footnote{Refs. \cite{Baier:2008js},\cite{Romatschke:2006bb} aimed to evalute momentum broadening in an anisotropic plasma. We believe they incorrectly assumed a KMS condition in the non-equilibirum setup, see e.g. Eq. (5) in \cite{Baier:2008js}. Doing so ignores the details of how soft gluons are emitted by an anisotropic distribution of quasiparticles.} This requires evaluating the \(rr\) correlator microscopically in a non-equilibrium plasma. We note that longitudinal momentum broadening is
\beq
\widehat{q}_L := \frac{d\langle (\Delta p_z)^2\rangle}{dt} =  \int d^4 Q\; q^2_{z}\frac{d\Gamma}{d^4 Q}
\eeq
for a parton travelling in the \(z\) direction
which can similarly be shown to be
\beq
\widehat{q}_L \approx  g^2 C_F \int \frac{d^4 Q}{(2\pi)^4}\; q_z^2 \,D_{rr}^{\mu\nu}(Q) v_{\mu} v_{\nu} \,2\pi\delta(v\cdot Q)
\eeq

In thermal equilibrium, many of our results simplify. The KMS relation gives that 
\beq
\label{Eq:KMS_Drr}
D^{\mu\nu}_{rr}(Q) = \left( \frac{1}{2} + f_{\mathrm{B}}(q^0)\right) \left[D^{\mu\nu}_{\ret}(Q) - D^{\mu\nu}_{\adv}(Q) \right].
\eeq
where \(f_{\mathrm{B}}(q^0) = 1/(e^{q^0/T}-1)\) is the Bose-Einstein distribution.
Using that \(f_{\mathrm{B}}(q^0) \approx T/q^0\) for soft gluons, the KMS relation leads to
\beq
\label{Eq:fluct_diss}
\widehat{q}_L = T \widehat{e}
\eeq
which is a fluctuation-dissipation relation linking momentum broadening due to fluctuating soft gluons in the medium and energy loss which happens through dissipation by the jet's radiation field.
Furthermore, in thermal equilibrium, the integral in Eq. \eqref{Eq:coll_kern} can be evaluated giving a simple, analytic result for the collision kernel \cite{Aurenche:2002pd,CaronHuot:2008ni},
\beq
\label{Eq:eq_coll_kern}
\mathcal{C}(\mathbf{q}_{\perp}) = g^2 C_F T \left(\frac{1}{\mathbf{q}_{\perp}^2} - \frac{1}{\mathbf{q}_{\perp}^2 + m_D^2} \right)
\eeq
where \(T\) is the temperature and \(m_D^2\) is the Debye mass.
We emphasize that Eqs. \eqref{Eq:KMS_Drr}, \eqref{Eq:fluct_diss} and \eqref{Eq:eq_coll_kern}  do not hold out of equilibrium and the \(rr\) propagator and the collision kernel \(\mathcal{C}\) need to be evaluated in detail.



\section{Density of soft gluons in an anisotropic medium}
\label{sec:Density of soft gluons}

The collision kernel in Eq. \eqref{Eq:coll_kern} for transverse momentum broadening depends on soft gluon density through the \(rr\) correlator. We will now derive the correator's symmetric component in an anisotropic medium.  The state of the medium is determined by the momentum distribution of hard quarks and gluons which are distributed anisotropically in momentum space. To fix ideas, we choose the momentum distribution introduced by Romatschke and Strickland \cite{Romatschke:2003ms},
\beq 
\label{Eq:RS_distr}
f(\mathbf{p}) = \sqrt{1+\xi}\, f_{\mathrm{eq}}\left(\sqrt{p^2 + \xi ( \mathbf{n} \cdot \mathbf{p})^2}\right).
\eeq
Here an equilibrium distribution \(f_{\mathrm{eq}}\), i.e. a Bose-Einstein distribution for gluons and a Fermi-Dirac distribution for quarks, is elongated or contracted in the direction of \(\mathbf{n}\) as quantified by the anisotropy parameter \(\xi\). We use \(\Lambda\) for the parameter corresponding to temperature in thermal equilibrium. The normalization factor \(\sqrt{1+\xi}\) guarantees that the number density of hard particles is the same as in thermal equilibrium. More general anisotropic distributions and the corresponding plasmons have been considered in the literature, see e.g. \cite{Carrington:2021bnk, Kasmaei:2018yrr}.

In momentum space the \(rr\) correlator is
\beq
\label{Eq:rr_eq}
D^{\mu\nu}_{rr}(Q) = D^{\mu\omega}_{\ret}(Q)\, \Pi^{\omega\xi}_{aa}(Q)\, D^{\xi\nu}_{\adv}(Q).
\eeq
where identical indices are contracted.\footnote{We use modern summation convention where \( A^{\mu} B^{\mu}  = A_{\mu} B^{\mu} =  g_{\mu\nu} A^{\mu} B^{\nu}\).} Here two soft gluon excitations are sourced with probability \(\epsilon_{\mu}\epsilon^*_{\nu}\Pi^{\mu\nu}_{aa}\) where \(\epsilon_{\mu}\) is the gluon polarization. The index \(aa\) on the self-energy comes from the \(r/a\) basis in the real-time formalism \cite{Ghiglieri:2020dpq}. The gluons then evolve in time as given by \(D_{\ret}\) and \(D_{\adv}\). This expression assumes a system initialized at time \(t_0=-\infty\) with the initial condition specified by quark and gluon momentum distributions. It furthermore assumes that the medium changes slowly enough to use Fourier transforms. We will discuss the validity of these assumption in Sec. \ref{sec:instabilities}.

 We assume that the system is in the hard thermal loops (HTL) regime. The hard quasiparticles at energy \(\Lambda\) then source soft gluons at energy \(g\Lambda\). These soft gluons have occupancy of order \(\sim 1/g\) and can be described with classical field theory. Their self-interaction is suppressed relative to interaction with hard quasiparticles \cite{Blaizot:2001nr}.
In this regime
\beq
\label{Eq:Pi_aa}
\begin{split}
\Pi^{\mu\nu}_{aa} &= g^2 \int \frac{d^3 p}{(2\pi)^3}\, v^{\mu} v^{\nu} 2\pi \delta (v\cdot Q) \bigg\rvert_{v=(1,\widehat{\mathbf{p}})} \\
&\times \left[ 2N_f f_q(\mathbf{p}) \left(1-f_q(\mathbf{p}) \right) + 2N_c f_g(\mathbf{p}) \left(1+f_g(\mathbf{p}) \right)\right]
\end{split}
\eeq
which gives the radiation of soft gluons by hard quarks, \(f_q(\mathbf{p})\), and hard gluons, \(f_g(\mathbf{p})\), including Bose enhancement and Pauli blocking \cite{Arnold:2002zm}.
As usually, \(N_f\) is the number of quark flavours and \(N_c\) is the number of colors.
The retarded correlator \(D^{\mu\nu}_{\ret}\) in  the HTL approximation in the Feynman gauge is
\beq
\label{Eq:retarded_corr}
D^{\mu\nu}_{\ret} =  i \left(\left[ P^2 - \Pi_{\ret}\right]^{-1}\right)^{\mu\nu}
\eeq
where 
\beq
\label{Eq:Pi_ret}
\Pi_{\ret}^{\mu\nu}(Q) = -g^2 \int \frac{d^3p}{(2\pi)^3} \, \frac{\partial f_{\mathrm{tot}}}{\partial P^{\omega}} \left[ - v^{\mu} g^{\omega \nu} + \frac{Q^{\omega}v^{\mu} v^{\nu}}{v\cdot Q + i \epsilon}\right] \bigg\rvert_{v=(1,\hat{\mathbf{p})}}
\eeq
with \(f_{\mathrm{tot}}(\mathbf{p}) = 2N_f f_q(\mathbf{p})  + 2N_c f_g(\mathbf{p})\) \cite{Mrowczynski:2000ed}.
It describes the propagation of soft gluons as they interact with hard particles in the medium. 
This retarded correlator  has been evaluated in \cite{Romatschke:2003ms} with the momentum distribution in Eq. \eqref{Eq:RS_distr}, see also \cite{Carrington:2014bla} for a detailed discussion. We will reproduce this derivation for completeness as our conventions differ. The advanced propagator is given by \(D_{\adv} = D_{\ret}^{\dagger}\). It is easy to see that \(D_{rr} \sim 1/g^3\).

Using  the ingredients we have assembled, we can evalute the \(rr\) correlator in an anisotropic plasma defined by the momentum distribution in Eq. \eqref{Eq:RS_distr}. This requires handling the tensor indices correctly. There are only four tensors in our medium: The metric \(g^{\mu\nu}\), the soft gluon momentum \(Q^{\mu} = (q^0,\mathbf{q})\), the fluid's velocity which we choose to be \(u^{\mu} = (1,\mathbf{0})\), as well as the direction of the anisotropy \(n^{\mu} = (0,\mathbf{n})\). It is convenient to define a new anisotropy vector as \(\tilde{n}^{\mu} = (0,\frac{\hat{\mathbf{n}}^i}{\sqrt{\hat{n}^2}})\) where 
\beq
\mathbf{\hat{n}} = \mathbf{n} - \frac{\mathbf{q}\cdot\mathbf{n}}{\mathbf{q}^2} \mathbf{q}.
\eeq
This guarantees that \(\tilde{n}^{\mu}\) is a unit spatial vector, orthogonal to both \(u^{\mu}\) and \(Q^{\mu}\). 

Using \(g^{\mu\nu}\), \(Q^{\mu}\), \(u^{\mu}\) and \(\tilde{n}^{\mu}\), we can construct seven symmetric, second-rank tensors. Additionally, both \(\Pi_{aa}\) and \(\Pi_{\ret}\) satisfy \(Q_{\mu} \Pi^{\mu\nu} = 0\) as a result of gauge invariance in the HTL approximation.
This requirement gives three independent equations since there are three vectors in \(\nu\). Therefore, the number of tensors we need to express the self-energies is reduced to four. A convenient choice for the first tensor is 
\beq
P_T^{ij} = \delta^{ij} - \frac{q^i q^j}{\mathbf{q}^2}
\eeq
which is transverse to the gluon momentum, 
with all other components zero. We choose the second tensor to be longitudinal to the gluon momentum,
\beq
P_L^{\mu\nu} = \frac{Q^{\mu}Q^{\nu}}{Q^2} - g^{\mu\nu} - P_T^{\mu\nu}
\eeq
Both of these tensors are present in thermal equilibrium \cite{Kapusta:2006pm}. 
The third tensor describes propagation along the anisotropy axis,
\beq 
C^{\mu\nu} = \tilde{n}^{\mu} \tilde{n}^{\nu}.
\eeq
Finally, the fourth tensor is given by 
\begin{align}
\begin{split}
D^{00} &= 0,\\
D^{0i} &= D^{i0} = \frac{\mathbf{q}^2}{q^0} \tilde{n}^i,\\
D^{ij} &= q^i \tilde{n}^j + q^j \tilde{n}^i
\end{split}
\end{align}
which mixes the anisotropy direction and the gluon three-momentum. The most convenient choice of the four tensors turns out to be \(P_T\), \(C\), \(D\) and \(E = P_T - C\) which we will use in what follows. An alternative definition of tensors is found in \cite{Dumitru:2007hy}.

Using this basis of tensors, the \(aa\) self-energy is
\beq 
\label{Eq:Piaa_decomp}
-i \Pi_{aa} = \alpha P_L + \beta E + \gamma C + \delta D,
\eeq
and it is straightforward to show that
\beq
\begin{split}
\alpha &= \frac{Q^2}{\mathbf{q}^2} \Pi^{00} \\
\gamma &= \Pi^{ij}\tilde{n}^{i} \tilde{n}^{j} \\
\delta &= \frac{q^0}{\mathbf{q}^2} \Pi^{0i} \tilde{n}^{i} \\
\beta &= -\alpha - \gamma - \Pi^{\mu}_{\mu}
\end{split}
\eeq
Explicit expressions for the components are collected in 
App. \ref{sec:Appendix_selfenergy}, with one integral left to be done numerically. Similarly, the retarded self-energy is 
\beq 
\label{Eq:Piret_decomp}
-i \Pi^{\mu\nu}_{\ret} = \Pi_L P_L^{\mu\nu} + \Pi_e E^{\mu\nu} + \Pi_c C^{\mu\nu} + \Pi_d D^{\mu\nu}
\eeq
with the components, which reproduce \cite{Romatschke:2003ms}, found in App. \ref{sec:Appendix_selfenergy}. 
Substituting in Eq. \eqref{Eq:retarded_corr} then gives the well-known result for the retarded propagator in Feynman gauge,
\beq
\begin{split}
D^{\mu\nu}_{\ret} &= \frac{-iQ^{\mu}Q^{\nu}}{\left(Q^2\right)^2} + iE^{\mu\nu}\tilde{D}^B_{\ret} \\
&+ i\left[ (Q^2-\Pi_c)P_L^{\mu\nu} + (Q^2-\Pi_L)C^{\mu\nu} + \Pi_d D^{\mu\nu}\right] \tilde{D}^A_{\ret}
\end{split}
\eeq
where the denominators \(\tilde{D}^A_{\ret}\) and \(\tilde{D}^B_{\ret}\) are defined in Eqs. \eqref{Eq:DretA} and \eqref{Eq:DretB}.

In order to derive the \(rr\) correlator, we must contract different tensors. This can conveniently be described by commutators and anticommutators. The anticommutator
\beq
\{X,Y\}^{\mu\nu} = X^{\mu\omega} \;Y_{\omega}^{\;\nu} + Y^{\mu\omega} \;X_{\omega}^{\;\nu},
\eeq
is guaranteed to be symmetric so our set of four tensors is closed under anticommutation. In fact, one can show that
\begin{align}
\begin{split}
\label{Eq:anticomm}
P_L^2 &= - P_L \\
E^2 &= -E \\
C^2 &= -C \\
D^2 &= -\frac{Q^2 \mathbf{q}^2}{q_0^2} \left(C+P_L \right) \\
\{E,P_L\} &= \{E,C\} = \{E,D\} = \{P_L,C\} = 0 \\
\{P_L,D\} &= \{C,D\} =  -D. \\
\end{split}
\end{align}
It can be shown that our choice of tensors makes the greatest number of anticommutators vanish, justifying the choice of \(E = P_T - C\). Nevertheless, the tensors do not form an orthogonal basis as some anticommutators of different tensors do not vanish. This means that the propagation of gluons mixes some of these modes.

 Momentum broadening depends on \(D_{rr}^{\mu\nu}\widehat{K}_{\mu} \widehat{K}_{\nu}\) so we only need the symmetric part,
\beq
\begin{split}
D^{\left(\mu\nu\right)}_{rr} &:= \frac{1}{2}\left[D_{rr} + D^{\dagger}_{rr}\right]^{\mu\nu} \\
&= \frac{1}{2} \left[ D_{\ret} \left( -i \Pi_{aa}\right) D_{\adv} + D_{\adv} \left( -i \Pi_{aa}\right) D_{\ret} \right]^{\mu\nu}
\end{split}
\eeq
Using the general relation 
\beq 
XYZ + ZYX = \frac{1}{2} \left[\{X,\{Y,Z\}\} - \{Y,\{Z,X\}\} + \{Z,\{X,Y\}\}\right]
\eeq
for tensors \(X\), \(Y\), \(Z\), we can evaluate the symmetric component of \(D_{rr}\) using the anticommutation relations in Eq. \eqref{Eq:anticomm}. The final result is that 
\begin{align}
\label{Eq:rr_total_aniso}
\begin{split}
D^{\left(\mu\nu\right)}_{rr} = & - \tilde{D}_{\ret}^A \left( \tilde{D}_{\ret}^A\right)^{*}  \\
\times \Bigg[ &\left\{\alpha \left|X\right|^2 - 2 \delta R \: \mathrm{Re}\! \left( X W^*\right)  + \gamma  R \left| W \right|^2 \right\} P_L^{\mu\nu}\\
+\;&\left\{ \gamma \left| Z\right|^2 - 2 \delta R \:\mathrm{Re}\! \left(Z W^* \right)  +\alpha R \left| W\right|^2 \right\} C^{\mu\nu}  \\
+ \; &\Big\{-\alpha\: \mathrm{Re}\! \left(X W^* \right) - \gamma\: \mathrm{Re}\! \left( Z W^*\right) \\
+\;&\delta\: \mathrm{Re}\! \left(X Z^* \right) + \delta R \left| W\right|^2  \Big\} D^{\mu\nu}\Bigg]\\
&- \tilde{D}_{\ret}^B \left( \tilde{D}_{\ret}^B\right)^{*} \beta E^{\mu\nu} \;\; \\
\end{split}
\end{align}
where
\beq
X = Q^2 - \Pi_c,
\eeq
\beq
Z = Q^2 - \Pi_L,
\eeq
\beq
W= - \Pi_d.
\eeq
and
\beq
\label{Eq:DretA}
\tilde{D}_{\ret}^{A} = \frac{1}{\left(Q^2 - \Pi_L \right) \left( Q^2 - \Pi_c\right) - R\: \Pi_d^2}
\eeq
and
\beq
\label{Eq:DretB}
\tilde{D}_{\ret}^{B} = \frac{1}{Q^2-\Pi_e}
\eeq
with \(R = Q^2 \mathbf{q}^2/\left(q^0\right)^2\). Previously, the \(00\) component of the \(rr\) correlator had been derived \cite{Nopoush:2017zbu}.



\section{Instabilities in an anisotropic plasma}
\label{sec:instabilities}


Having calculated the density of soft gluons in an anisotropic medium,  Eq. \eqref{Eq:rr_total_aniso}, we expect to obtain the rate of transverse momentum broadening from Eq. \eqref{Eq:coll_kern}. However, this gives a divergent collision kernel \(\mathcal{C}(\mathbf{q}_{\perp})\), see App. \ref{sec:Appendix_cutoff} for the mathematical details. The divergence is due to instabilities that are always present in an anisotropic plasma.
The physical origin of these instabilities is widely discussed in the literature, see \cite{Mrowczynski:1988dz,Mrowczynski:1993qm}
for an early discussion, as well as \cite{Arnold:2003rq, Hauksson:2020etn}, and \cite{Mrowczynski:2016etf} for a review. In essence, hard quasiparticles spontaneously break into filaments of currents which source soft chromomagnetic fields. These soft fields deflect the hard quasiparticles which makes the currents even stronger. This sources even stronger soft gluon fields, and so on, leading to an exponential growth in the density of soft gluons.


The divergence in the collision kernel forces us to examine implicit assumptions we made in Sec. \ref{sec:Density of soft gluons}. Firstly, we assumed a slowly changing medium. This means that the momentum distribution of hard quasiparticles, which appear in bare propagators, remains constant during jet momentum broadening. Secondly, we specified these momentum distribution at time \(t_0 = -\infty\). This made Fourier transforms possible, leading to simple expressions such as the self-energies in Eqs. \eqref{Eq:Pi_aa} and \eqref{Eq:Pi_ret} in momentum space.

Specifying an initial condition at time \(t_0 = -\infty\) might seem justified in a slowly changing medium.
However, this assumption is invalidated by instabilities. 
In a strict HTL regime, the density of soft gluons continues to grow from the initial time due to instabilities. Thus at the time of a gluon kick on a jet parton, the density of gluons will have grown to be infinite and the rate of momentum broadening diverges.\footnote{
This problem of a divergent rate due to instabilities is not unique to momentum broadening. All probes that depend on the density of soft gluons suffer from the same spurious divergences in an anisotropic medium with initial condition specified at time \(t_0 = -\infty\). This includes the rate of photon production through bremsstrahlung, medium-induced jet splitting and medium-induced quasiparticle splitting \cite{Arnold:2002zm} which all depend on momentum broadening. Furthermore, the imaginary part of the heavy quark potential also diverges under these assumptions \cite{Nopoush:2017zbu}.}

The solution to this problem is to specify the initial momentum distribution at a finite time, \(t_0 = 0\), and to take into account its time evolution. Our recent work \cite{Hauksson:2020wsm} did this analytically by deriving the evolution of the \(rr\) correlator in an anisotropic system. We assumed a small anisotropy in the momentum distribution of hard particles, \(\xi^{3/2} \ll g^2\) ,
where
\beq
\xi \sim \frac{\left| \langle p_z\rangle - \langle p_{\perp} \rangle \right|}{\langle p_z\rangle}.
\eeq 
This small anisotropy is needed for an adiabatic approximation. Specifically, the growth rate of instabilities is \(\gamma \sim \xi^{3/2} g \Lambda\) \cite{Kurkela:2011ti} and we need the growth rate to be slower than the process of interest. 

Assuming a small anisotropy, \(\xi \ll 1\) the retarded correlator is 
\beq
\label{Eq:ret_final}
G_{\ret}(t_x,t_y; \mathbf{p}) = \int_{\alpha} \frac{dp^0}{2\pi} \;e^{-ip^0(t_x-t_y)} G_{\ret}(p^0,\mathbf{p}).
\eeq
shortly after initialization in our setup when the HTL approximation is still valid.
Here \(G_{\ret}(p^0,\mathbf{p}) = \left[(G_{\ret}^0(P))^{-1}- \Pi_{\ret}(P)\right]^{-1}\) where \(\alpha\) is a contour that goes above all poles, see \cite{Hauksson:2020wsm}, and the self-energy is given by Eq. \eqref{Eq:Pi_ret}. In particular it goes above poles \(\omega = i\gamma\) with \(\gamma > 0\) that correspond to instability modes and lie in the upper half  of the complex plane in \(\omega\)  \cite{Romatschke:2003ms}. Such an instability pole gives exponential growth in the time domain where  \(G_{\ret}(t_x,t_y) \sim \theta(t_x - t_y)\,e^{\gamma(t_x-t_y)}\). 

Some earlier work on energy loss in an anisotropic plasma \cite{Romatschke:2003vc, Romatschke:2004au} implicitly assumed an integration contour along the real axis. This gives an incorrect contribution \(\theta(t_y-t_x) e^{\gamma(t_x - t_y)}\) to the retarded propagator in the time domain and leads to incorrect results for energy loss. We discuss this further in Appendix \ref{sec:Appendix_energyloss}.





In \cite{Hauksson:2020wsm} we furthermore used a separation of scales to evaluate the \(rr\) correlator in a system initialized at time \(t_0 = 0\). We write the retarded correlator in momentum space as
\beq
\label{Eq:ret_scale_sep}
G_{\ret}(K) = \widehat{G}_{\ret}(K) + \sum_{i} \frac{A_i}{k^0 - i\gamma_i}
\eeq
where \(i \gamma_i\) are all poles of order \(\xi^{3/2} g \Lambda\) including instability poles and \(\widehat{G}_{\ret}\) describes modes of order \(g\Lambda\).\footnote{The power in \(\xi^{3/2}g\Lambda\) is for example derived in \cite{Kurkela:2011ti}. More generally, we are separating into modes with frequency \(\sim g\Lambda\) and modes with frequency \(\ll g\Lambda\).} Then 
\beq
\begin{split}
\label{Eq:rr_final}
G_{rr}(t_x,t_y;\mathbf{k}) \approx &\int \frac{dk^0}{2\pi}\;e^{-ik^0(t_x-t_y)} \widehat{G}_{rr}(K) \\ 
&\hspace{-2cm}+ \sum_{i,j} \;\frac{A_i \Pi_{aa}(0) A_j^*}{\gamma_i + \gamma_j} \left[ e^{\gamma_i \,t_x} e^{\gamma_j \,t_y}-1\right]
\end{split}
\eeq
shortly after initialization in our setup where
\beq
\widehat{G}_{rr}(K) = \widehat{G}_{\ret}(K) \,\Pi_{aa}(K)\, \widehat{G}_{\adv}(K) 
\eeq
and \(\widehat{G}_{\adv} = \widehat{G}_{\ret}^{*}\).
Modes of order \(g\Lambda\) are described by \(\widehat{G}_{rr}\). These fluctuating soft modes are continuously sourced by hard particles in the medium and they only depend on the particles' instantaneous momentum distribution \(f(\mathbf{p})\), given that the HTL approximation is valid. Furthermore, their density does not depend on how much time has passed since initialization. Instability modes at energy \(\xi g\Lambda\) are described  by the second term. They grow exponentially in time shortly after initialization.\footnote{Further discussion can be found in App. \ref{sec:App_correction}. In particular, we correct some wrong sign in a heuristic discussion of the \(rr\) propagator in an unstable plasma in our earlier paper \cite{Hauksson:2020wsm}. These sign mistakes in no way change the results of that paper.}

Eq. \eqref{Eq:rr_final} for the \(rr\) correlator in an anisotropic plasma is only valid for the first instants after a system is initialized. It corresponds to the early times  in classical-statistical simulations of heavy-ion collisions \cite{Berges:2013eia,Berges:2013fga}. However, we are interested in the collision kernel at later stages of heavy-ion collisions where instability modes have been saturated and the system has achieved a HTL-like scale separation  after going through a non-thermal fixed point \cite{Berges:2014bba}. During these later stages, namely the kinetic theory stage and hydrodynamics stage, the part of \eqref{Eq:rr_final} describing fluctuating modes remains valid: Those gluons are continuosly sourced by the hard quasiparticles in the medium and their density does not depend on the history of the medium. On the other hand, the ultrasoft instability modes at scale \(\xi^{3/2} g\Lambda\) have evolved through self-interaction which is not captured by our analytic calculation. 

In this paper, we focus on momentum broadening due to the fluctuating modes.  More precisely, we choose a scale \(\omega_{\mathrm{cut}}\) separating the fluctuating modes at energy \(g\Lambda\) and the instability modes at energy \(\xi^{3/2} g\Lambda\) with
\beq
\xi^{3/2} g \Lambda \ll \omega_{\mathrm{cut}} \ll g\Lambda.
\eeq
We will focus on modes  at \(\omega > \omega_{\mathrm{cut}}\) which are captured by the HTL \(rr\) correlator in Eq. \eqref{Eq:rr_total_aniso}. Conversely, we will not include the effect of ultrasoft modes, including instability modes, at \(\omega < \omega_{\mathrm{cut}}\), as their dispersion relation in heavy-ion collisions is not captured by our analytic calculation. These ultrasoft modes only occupy a small portion of momentum space and should give a limited contribution to momentum broadening, except for potential divergences. We explore this further below.

\begin{figure}
    \centering
    \includegraphics{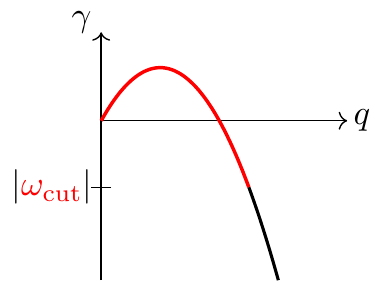}
    \caption{Definition of \(\omega_{\mathrm{cut}}\). Instability poles \(i\gamma\) are plotted with momentum \(q\) for a particular direction. All poles in red are subtracted, i.e. all poles in the upper half plane, as well as all poles in the lower half plane with \(|\gamma| < \omega_{\mathrm{cut}}\)}
    \label{Fig:inst_subtr}
\end{figure}

In practice, we locate instability poles numerically and subtract their contribution from the retarded and advanced correlators. These poles are only present for small momenta, \(q \sim \xi^{1/2} g\Lambda\). We subtract all instability poles in the upper half complex plane which correspond to exponential growth and which should not be treated in momentum space. We furthermore subtract poles in the lower half plane with \(\omega = i\gamma\) and \(|\gamma| < \omega_{\mathrm{cut}}\), see Fig. \ref{Fig:inst_subtr}. 
These modes are on the second Riemann sheet in \cite{Romatschke:2004jh} which appears when the branch cut corresponding to Landau damping is modified \cite{Hauksson:2020wsm}.\footnote{Modifying the branch cut does not change the correlator in the time domain where it is properly defined, see \cite{Kurkela:2017xis}.}
This gives an \(rr\) correlator for fluctuating modes 
\beq
\widehat{G}_{rr}(K) = \widehat{G}_{\ret}(K) \,\Pi_{aa}(K)\, \widehat{G}_{\adv}(K).
\eeq
which coincides with that of Eq. \eqref{Eq:rr_total_aniso} except that ultrasoft instability poles are subtracted in the retarded and advanced correlator. For numerical convenience we perform the subtraction of a pole \(\omega = i\gamma\) in \(\tilde{D}_{\ret}^A\) and \(\tilde{D}_{ret}^B\) from Eq. \eqref{Eq:DretA} and \eqref{Eq:DretB}. Defining \(\tilde{D}_{ret}^A = 1/A(\omega)\), we write exactly
\beq
\begin{split}
\frac{1}{A(\omega)} &= \frac{1}{\frac{A(\omega)}{(\omega-i\gamma)}(\omega-i\gamma)} \\ 
 &= \frac{1}{A(\omega) - \left( \frac{A(\omega)}{\omega - i\gamma}\right)^2}  - \frac{1}{(\omega - i\gamma) \,-\, \frac{A(\omega)}{\omega - i\gamma}} \frac{1}{\omega - i\gamma}
\end{split}
\eeq
and subtract the second term which has a pole at \(\omega = i\gamma\) and the correct residue as \(\omega \approx i\gamma\).  The first term has no pole at \(\omega = i\gamma\) because
\beq
A(\omega) - \left( \frac{A(\omega)}{\omega - i\gamma}\right)^2 = \frac{A(\omega)}{\omega - i\gamma} \left[\omega -i\gamma - \frac{A(\omega)}{\omega - i\gamma}\right]
\eeq
where \(\frac{A(\omega)}{\omega - i\gamma}\) is finite everywhere and non-zero at \(\omega = i\gamma\).


\section{Results}
\label{sec:Results}

\subsection{Dependence on \(\omega_{\mathrm{cut}}\)}

\begin{figure}
    \centering
    \includegraphics[width=0.17\textwidth]{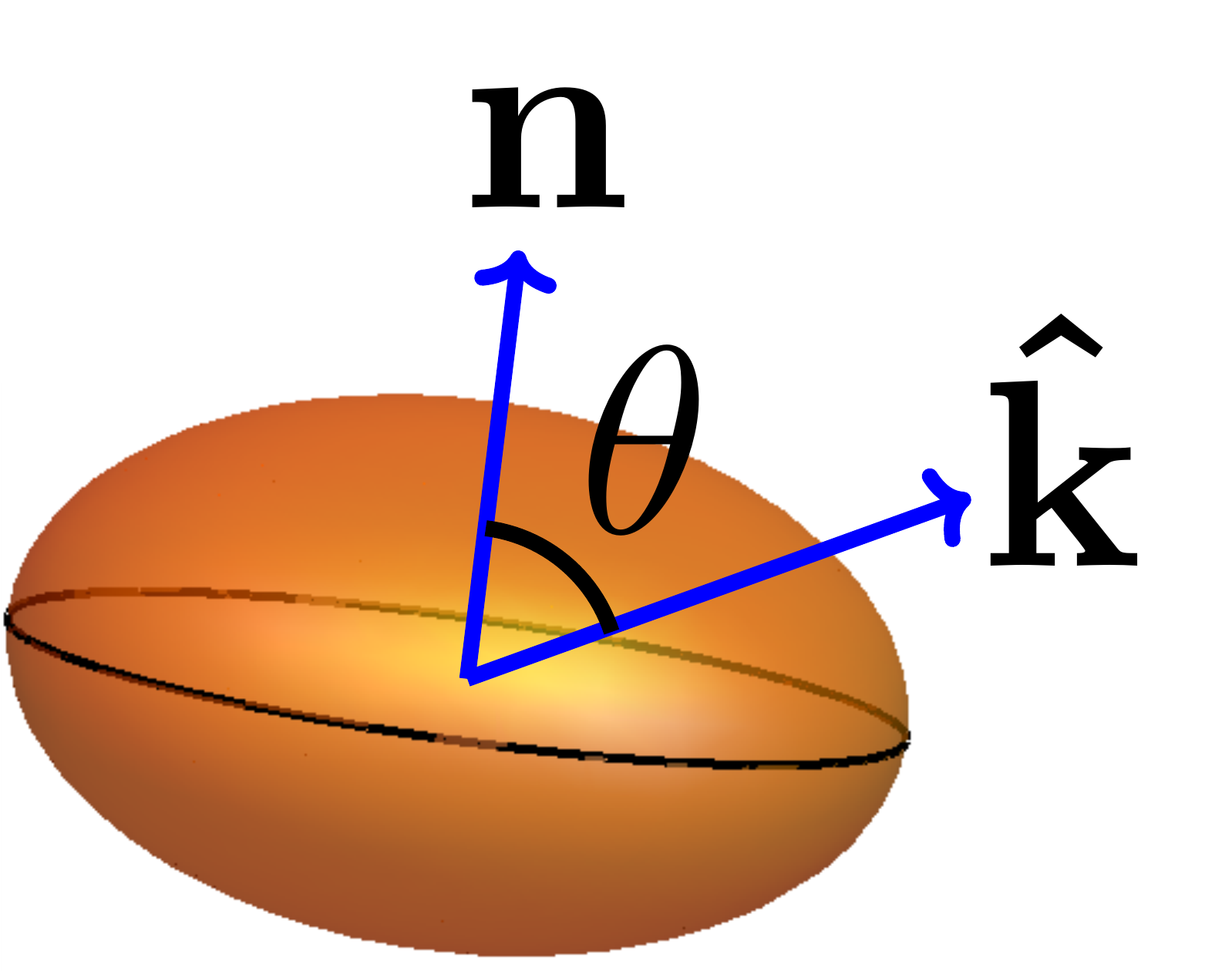}
    \caption{The jet parton direction is defined through \(\theta\), the angle between the jet momentum \(\hat{\mathbf{k}}\) and the anisotropy vector \(\mathbf{n}\). Pictured is the case of \(\xi>0\) where the momentum distribution is oblate. For \(\xi<0\) the distribution is prolate.}
    \label{fig:def_of_theta}
\end{figure}

\begin{figure}
    \centering
    \begin{subfigure}[b]{0.45\textwidth}
         \centering
         \includegraphics[width=\textwidth]{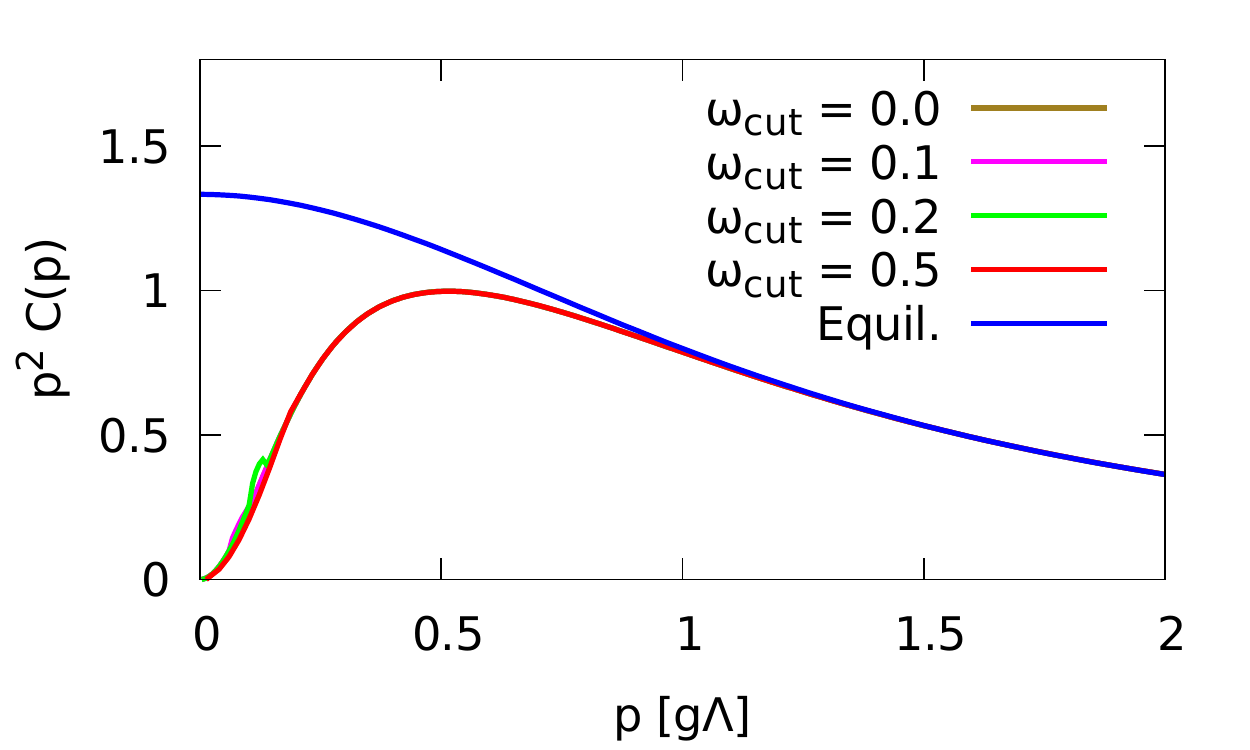}
         \caption{Anisotropy \(\xi\) = -0.1, jet direction \(\theta = \pi/2\), angle in transverse plane \(\phi=0\)}
         \label{Fig:}
     \end{subfigure}
     \hfill
    \begin{subfigure}[b]{0.45\textwidth}
         \centering
         \includegraphics[width=\textwidth]{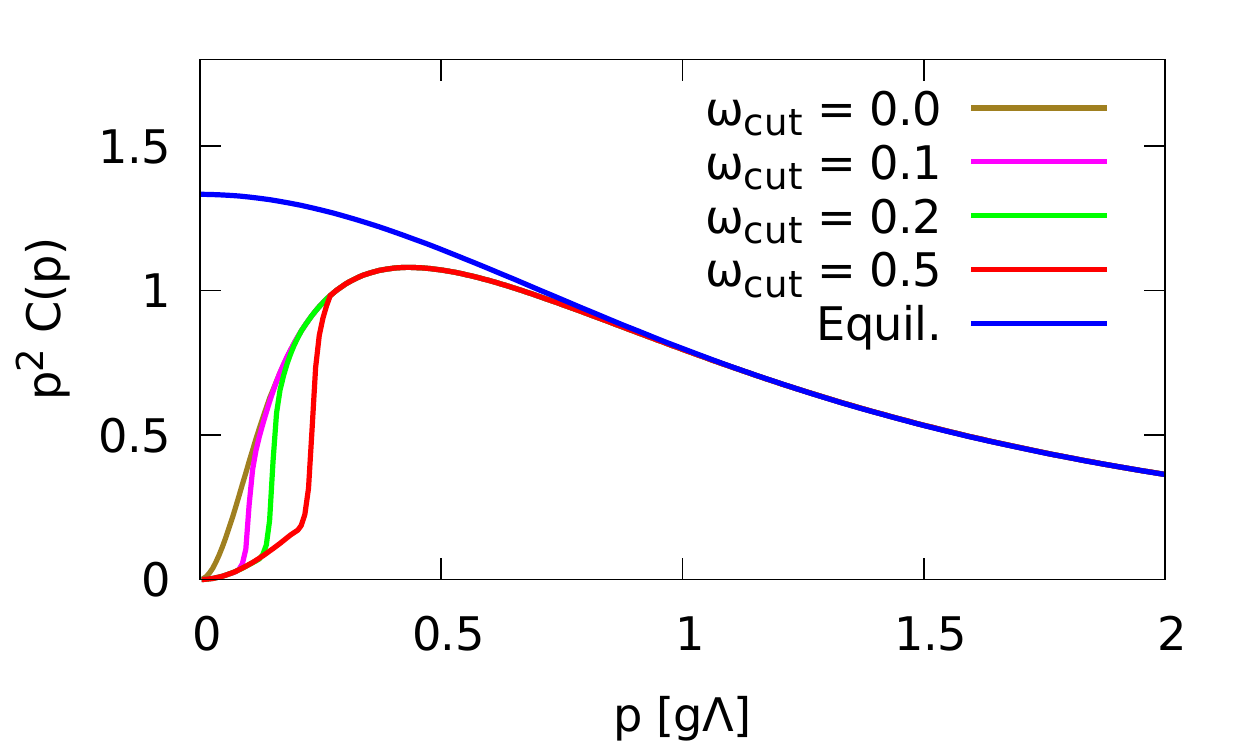}
         \caption{\(\xi\) = -0.1, \(\theta = \pi/3\), \(\phi=0\)}
         \label{Fig:}
     \end{subfigure} 
    \caption{
    The amount of transverse momentum broadening \(\mathbf{p}_{\perp}^2 \mathcal{C}(\mathbf{p})\) at momentum \(\mathbf{p}_{\perp}\) in units of \(g^2 \Lambda\). The transverse momentum \(\mathbf{p}_{\perp}\) is measured in units of \(g\Lambda\). The different curves have different \(\omega_{\mathrm{cut}}\), the cut between fluctuating modes and ultrasoft instability modes which are subtracted. The equilibrium result with temperature \(T = \Lambda\) and \(\xi =0\) is shown for comparison. The dependence on the cut is mild for these values of \(\xi \) and \(\theta\). A similar picture holds in general for \(\xi <0\) at \(\pi/4 < \theta \leq \pi/2\), and for \(\xi >0\) at \(0\leq \theta < \pi/4\).
    }
\label{Fig:cut_nice}
\end{figure}

The collision kernel \(\mathcal{C}(\mathbf{p}_{\perp})\) depends on both medium properties and the direction of the jet. Specifically, the kernel depends on the anisotropy \(\xi\) of the medium, the jet direction \(\theta\), the magnitude of the transverse momentum kick \(p_{\perp}\) and its direction \(\phi\). The medium anisotropy \(\xi\) is defined in Eq. \eqref{Eq:RS_distr} for the momentum distribution of quasiparticles. It describes elongation or contraction of an isotropic distribution in a direction \(\mathbf{n}\). The direction of the jet is specified by \(\theta\), the angle between the jet momentum and \(\mathbf{n}\), see Fig. \ref{fig:def_of_theta}. The symmetry of our setup means that only values of \(0\leq \theta \leq \pi/2\) need to be considered. The direction of a momentum kick in the plane transverse to the jet momentum is given by an angle \(\phi\). We choose \(\phi=0\) when \(\mathbf{p}_{\perp}\) is in the plane defined by the jet direction and \(\mathbf{n}\).  

\begin{figure}
    \centering
    \begin{subfigure}[b]{0.45\textwidth}
         \centering
         \includegraphics[width=\textwidth]{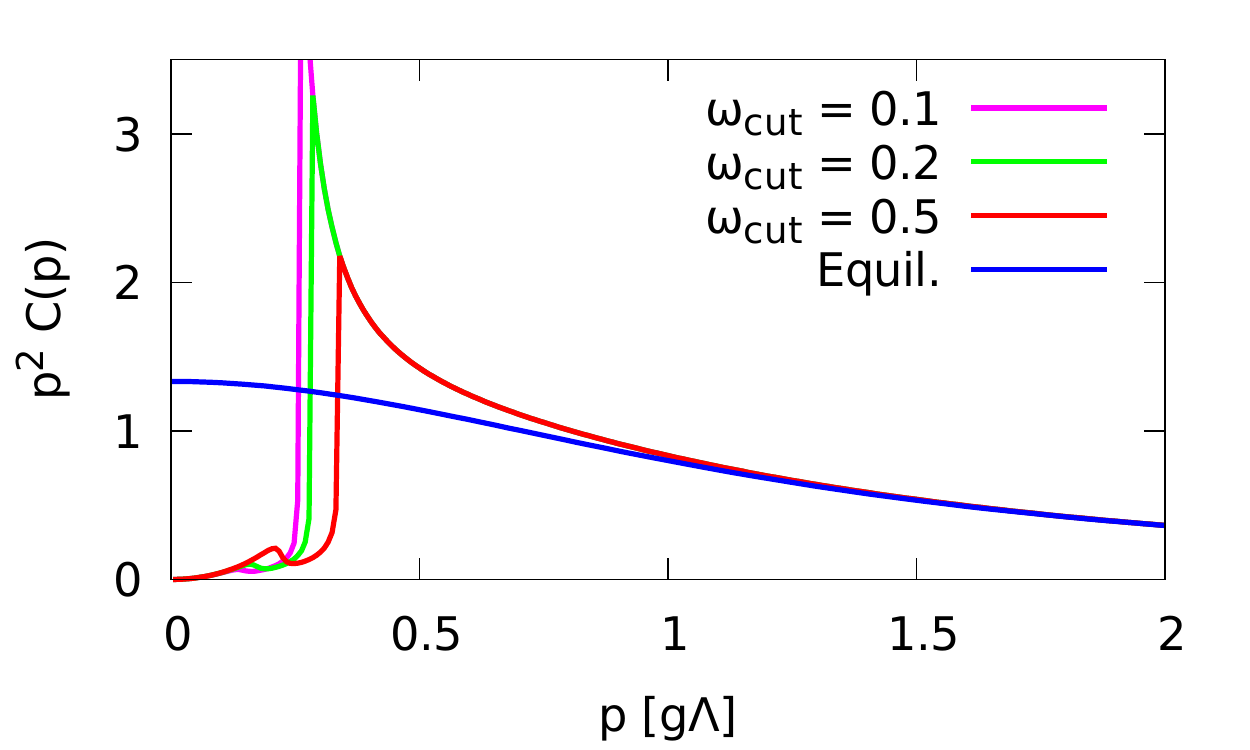}
         \caption{\(\xi\) = -0.1, \(\theta = 0\), \(\phi=0\)}
         \label{Fig:}
     \end{subfigure}
     \hfill
    \begin{subfigure}[b]{0.45\textwidth}
         \centering
         \includegraphics[width=\textwidth]{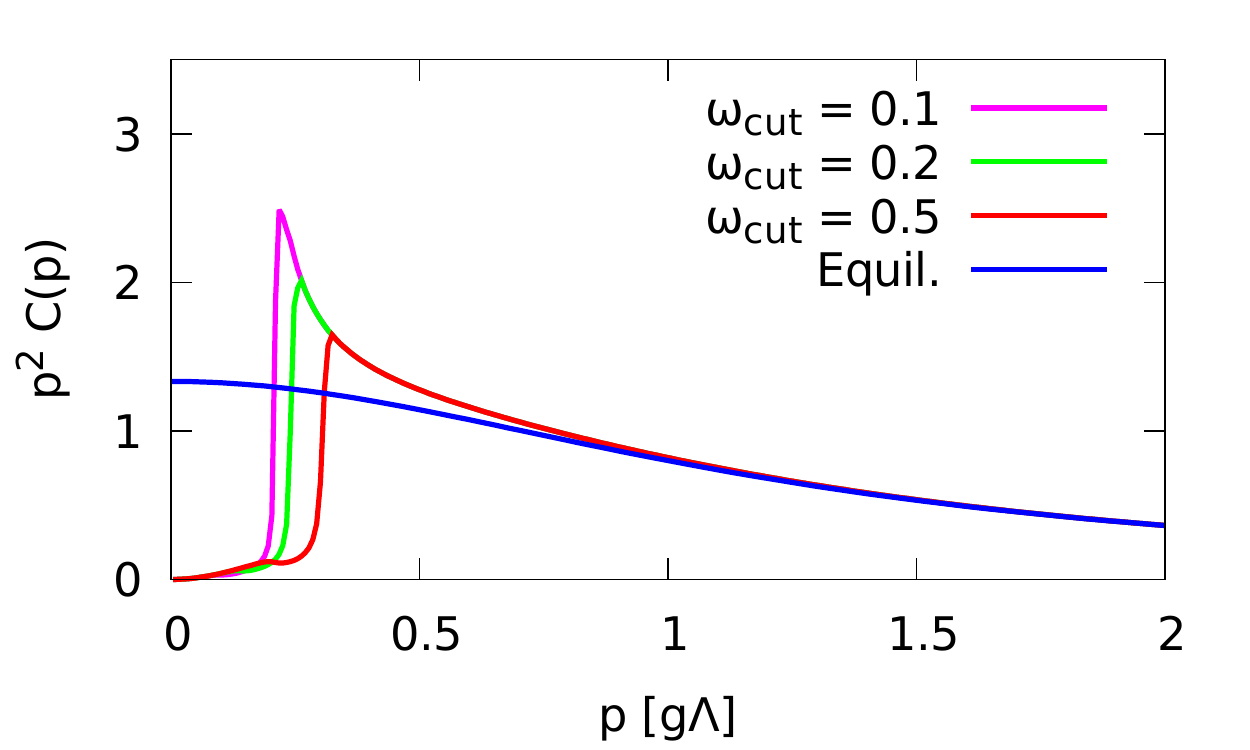}
         \caption{\(\xi\) = -0.1, \(\theta = \pi/6\), \(\phi=0\)}
         \label{Fig:}
     \end{subfigure} 
    \caption{ The dependence of transverse momentum broadening on \(\omega_{\mathrm{cut}}\) for two values of \(\theta\). 
    The dependence on the cut is more pronounced than in Fig. \ref{Fig:cut_nice} due to slowly decaying modes transverse to the jet direction. A similar picture holds in general for \(\xi <0\) at \(0\leq \theta < \pi/4\) and for \(\xi >0\) at \(\pi/4 \leq \theta \leq \pi/2\).}
\label{Fig:cut_div}
\end{figure}

We have obtained consistent expressions for the collision kernel by subtracting ultrasoft instability poles at frequency below a cut \(\omega_{\mathrm{cut}}\), leaving fluctuating modes at energy \(\omega > \omega_{\mathrm{cut}}\) which are the focus of this study. It is important to explore how sensitive our results are to the exact value of the cut.\footnote{To subtract instability poles in the lower half plane on the second Riemann sheet, one needs the analytic continuation of the propagator to that Riemann sheet. The self-energies in Eq. \eqref{Eq:Piret_decomp} have one remaining integral and thus cannot be analytically continued in any simple way. Instead, we will use the expansion at low \(\xi\) for \(\Pi_{\ret}\)  derived in \cite{Romatschke:2003ms} and \cite{Carrington:2014bla} when estimating the effect of \(\omega_{\mathrm{cut}}\) in Figs. \ref{Fig:cut_nice} and \ref{Fig:cut_div}. As we subsequently choose \(\omega_{\mathrm{cut}} = 0.0\), with no poles in the lower half plane subtracted, we use the full results valid at all \(\xi\) in all other figures.} This measures the robustness of our results and the importance of ultrasoft modes in momentum broadening. We will write \(\omega_{\mathrm{cut}} = a_{\mathrm{cut}}\, \xi^{3/2} g \Lambda\) where \(a_{\mathrm{cut}}\) is a number.\footnote{For comparison the maximal growth rate from instabilities in \(\tilde{D}_{\ret}^A\) in Eq. \eqref{Eq:DretA} is \(\gamma_{max}  \approx 0.15 g\Lambda \xi^{3/2}\) \cite{Carrington:2014bla}. Thus a reasonable choice for the cut in the lower half plane is \(a_{\mathrm{cut}} \sim 0.1 - 0.5\).} All our results will assume a QCD plasma with three flavours of massless quarks.

The sensitivity of momentum broadening on the cut \(\omega_{\mathrm{cut}}\) is moderate and depends on the direction of the jet.: in Figs. \ref{Fig:cut_nice} and \ref{Fig:cut_div} we show the amount of transverse broadening at momentum \(\mathbf{p}_{\perp}\) as the cut is varied.\footnote{Subtracting poles with \(|\gamma| < \omega_{\mathrm{cut}}\) introduces kinks in \(\mathcal{C}(\mathbf{p}_{\perp})\), see Fig. \ref{Fig:cut_div}. Including the physics of ultrasoft modes would give a continuous description and remove these kinks.}  These figures are representative for the results at a negative anisotropy, \(\xi <0\). For jet direction \(\pi/4 \leq \theta \leq \pi/2\) as in Fig. \ref{Fig:cut_nice}, there is limited dependence on the cut and one gets consistent result by subtracting the instability poles. For  \(0\leq \theta < \pi/4\) as in Fig. \ref{Fig:cut_div} there is more dependence on the the value of the cut. 

The different qualitative behaviour in Figs. \ref{Fig:cut_nice} and \ref{Fig:cut_div} can be understood by analyzing the structure of instability poles. For negative anisotropy, when \(\pi/4 \leq \theta \leq \pi/2\) there are instability poles in \(D_{\ret}^A\) with momentum transverse to the direction of the jet \cite{Carrington:2014bla}.
Even after subtracting the poles below \(\omega_{\mathrm{cut}}\) there remain very slowly decaying modes with \(\omega = -i\gamma\), \(\gamma \gtrsim \omega_{\mathrm{cut}}\) which have momentum nearly transverse to the jet direction. Due to the slow decay of these modes, each excitation can impart transverse momentum to the jet for a long time.
Thus these modes can impart a great deal of transverse momentum which explains the rapid rise in Fig. \ref{Fig:cut_div}. As there are no such slowly decaying mode transverse to the jet direction for \(0\leq \theta < \pi/4\) one gets much less sensitivity to the cut.
The picture is opposite for positive anisotropy. Due to the location of instability poles, one gets little sensitivity to the cut for \(0\leq \theta < \pi/4\) like in Fig. \ref{Fig:cut_nice} and more sensitivity for \(\pi/4 \leq \theta \leq \pi/2\) like in Fig. \ref{Fig:cut_div}.

The dependence of \(\widehat{q}\) on the frequency cut can be estimated analytically, see App. \ref{sec:Appendix_cutoff}. One gets that 
\beq
\widehat{q} \sim g^4 \Lambda ^3 \xi^{3/2} \log \left( \sqrt{\xi} \,\omega_{\mathrm{cut}} \right).
\eeq
This relatively mild dependence is because of the small phase space in which one finds instabilities.

For the rest of this paper we will focus on values of \(\xi\) and \(\theta\) where there is less sensitivity to the cut, like in Fig. \ref{Fig:cut_nice}. For simplicity we will choose \(\omega_{\mathrm{cut}} = 0.0\). For other values of \(\xi\) and \(\theta\) like in Fig. \ref{Fig:cut_div}, the collision kernel is sensitive to the exact dispersion relation and occupation density of ultrasoft modes. Thus our calculation should eventually be complemented by detailed information on the far infrared in heavy-ion collisions. This would of course cancel any dependence on \(\omega_{\mathrm{cut}}\) in our calculation.

\subsection{Dependence on anistropy and jet direction}

\begin{figure}
    \centering
         \includegraphics[width=0.45\textwidth]{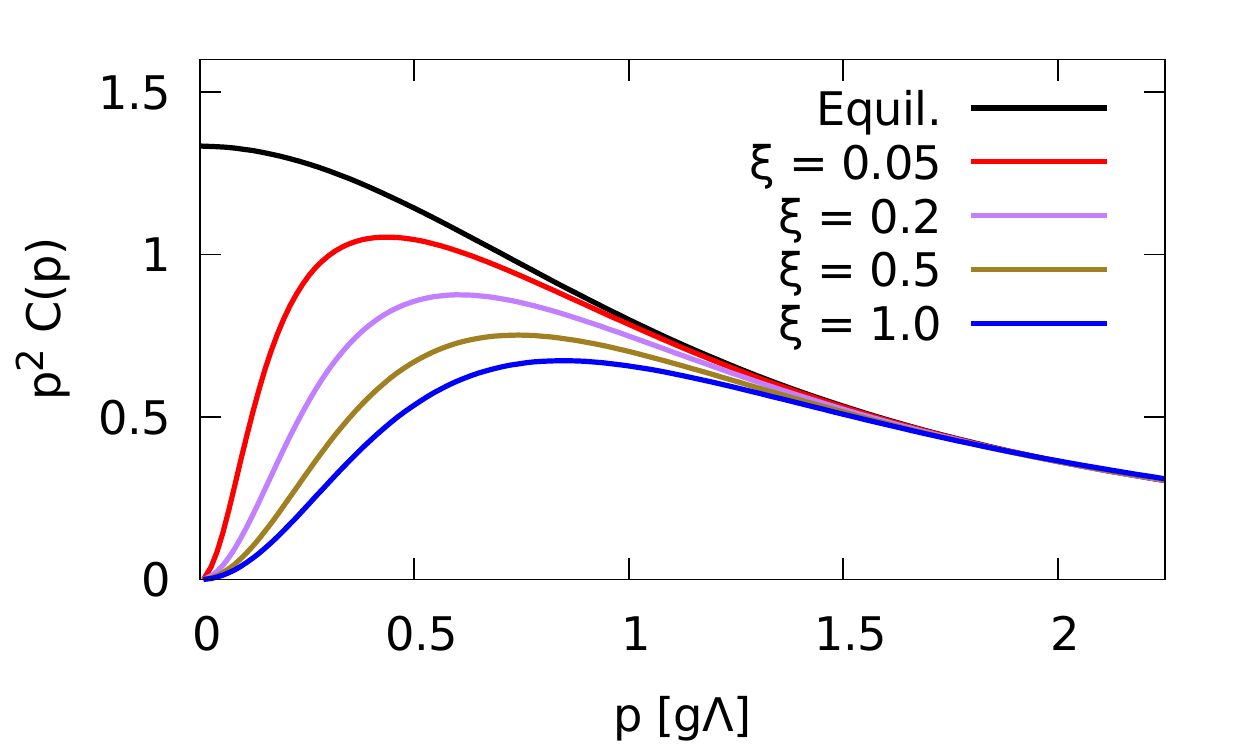}
         \caption{Transverse momentum broadening for a jet parton parallel to the anisotropy vector, \(\theta = 0\), for positive anisotropy. Momentum broadening is reduced as the anisotropy increases, especially for low and intermediate values of \(\mathbf{p}_{\perp}\). }
         \label{Fig:dep_on_xi_positive}
\end{figure}

\begin{figure}
    \centering
         \includegraphics[width=0.45\textwidth]{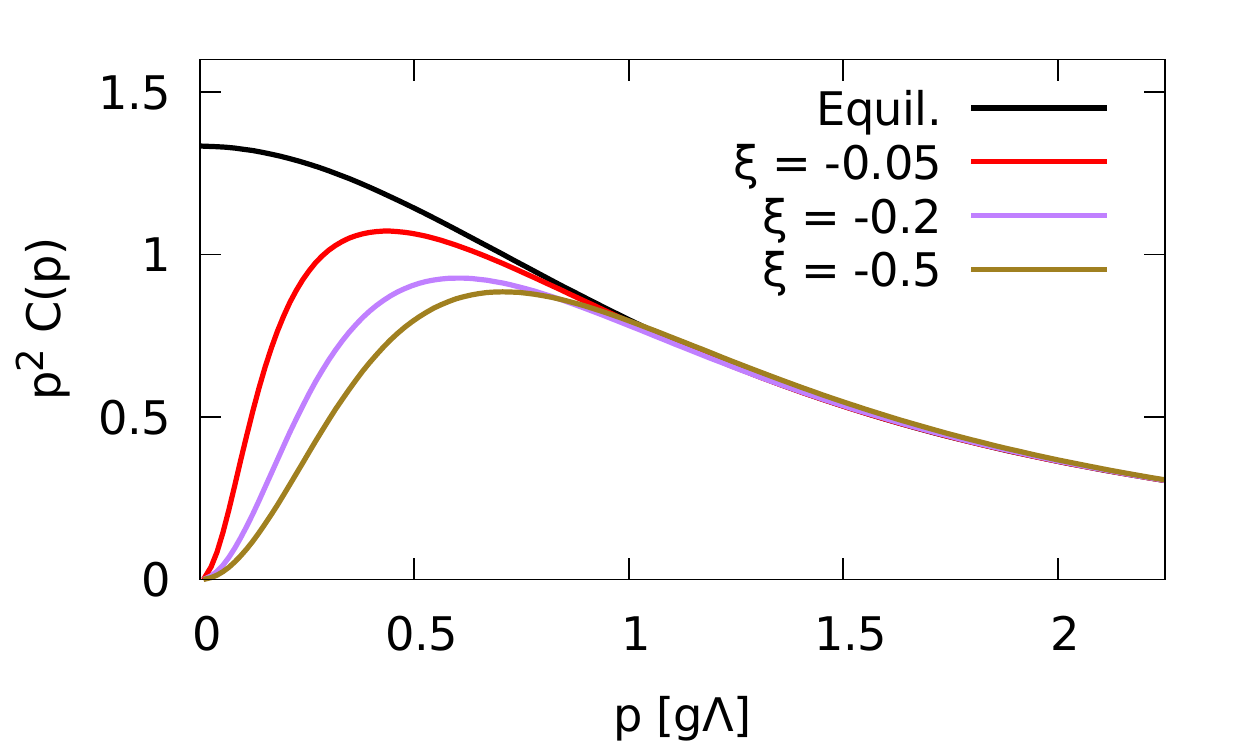}
         \caption{Transverse momentum broadening for a jet parton transverse to the anisotropy vector, \(\theta = \pi/2\), for negative anisotropy. As in Fig. \ref{Fig:dep_on_xi_positive}, momentum broadening is reduced in a more anisotropy medium. Here \(\phi = 0\).}
         \label{Fig:dep_on_xi_negative}
\end{figure}

Medium anisotropy reduces momentum broadening because medium screening is increased. In Fig. \ref{Fig:dep_on_xi_positive} momentum broadening is shown for different positive values of \(\xi\) where the jet parton is parallel to the anisotropy vector, \(\theta = 0\). In Fig. \ref{Fig:dep_on_xi_negative} momentum broadening is shown for negative values of \(\xi\) with the jet transverse to the anisotropy vector, \(\theta = \pi/2\). Both results are qualitatively similar. Deviation from equilibrium universally results in less momentum broadening.\footnote{This is universally true for values of \(\xi\) and \(\theta\) like in Fig. \ref{Fig:cut_nice} which have little sensitivity to \(\omega_{\mathrm{cut}}\). For values of \(\xi\) and \(\theta\) where information on ultrasoft modes is required like in Fig. \ref{Fig:cut_div}, it is not clear to us whether the same reduction in momentum broadening will take place.} This reduction is clear even at small anisotropy. The deviations from equilibrium are most pronounced at low and intermediate values of \(\mathbf{p}_{\perp}\) as at high values medium screening is unimportant. We have also checked the dependence on the angle \(\theta\) between the jet parton and the anisotropy vector. It is found to be fairly mild.


The qualitative difference between equilibrium and non-equilibrium momentum broadening is due to additional screening in the non-equilibrium plasma. The equilibrium collision kernel is \(\mathcal{C} \sim 1/\mathbf{q}_{\perp}^2\) for low transverse momentum as can be seen in Eq. \eqref{Eq:eq_coll_kern}. This is because of the absence of static screening for magnetic modes in a quark-gluon plasma in thermal equilibrium \cite{Aurenche:2002pd} 
while the electric mode gets screening from the Debye mass \(m_D^2\). Conversely, our non-equilibrium collision kernel goes to a constant at very low \(\mathbf{q}_{\perp}\). This is a sign of additional screening in a non-equilibrium medium. In particular, the self-energy component \(\Pi_c\) goes to zero in thermal equilibrium in the limit \(\omega \rightarrow 0\) denoting absence of magnetic screening but remains finite in the anisotropic case. This is true even for moderate values of the instability cutoff \(\omega_{\mathrm{cut}}\).  

\begin{figure}
    \centering
         \includegraphics[width=0.45\textwidth]{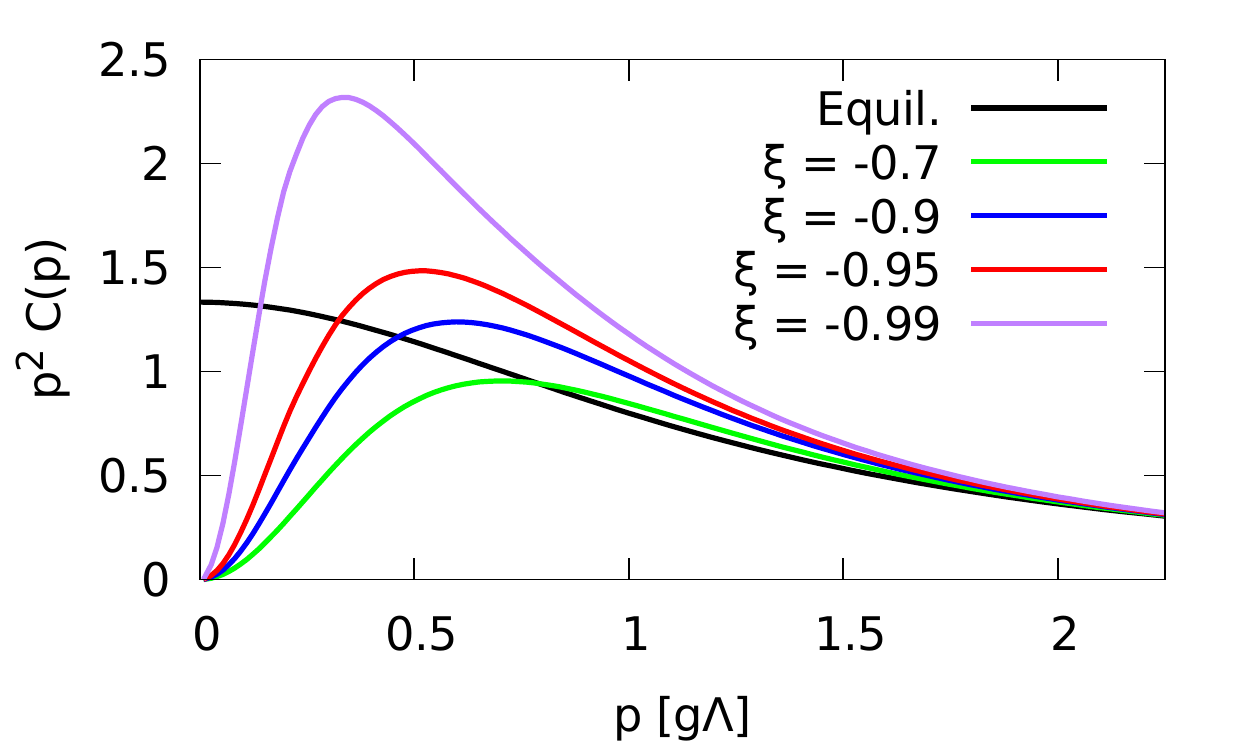}
         \caption{Transverse momentum broadening at high negative values of \(\xi\) with \(\theta = \pi/2\) and \(\phi = 0\). An additional peak develops at \(p_{\perp} \sim 0.5 g\Lambda\) relative to low and intermediate values of \(\xi\) as in Fig. \ref{Fig:dep_on_xi_negative}. This regime of extreme anisotropy stretches our assumption that modes can be separated into instability modes and anisotropy modes.}
         \label{Fig:dep_on_highxi_negative}
\end{figure}

Our separation of physics into instability modes and fluctuating modes makes most sense for low and intermediate values of \(\xi\). It is nevertheless interesting to see how the collision kernel behaves at extreme anisotropy in our setup. In Fig. \ref{Fig:dep_on_highxi_negative} we show the kernel for very large negative values of \(\xi\) which must be bounded from below by \(-1\). At extreme anisotropy, the collision kernel devolopes a pronounced peak around \(p_{\perp} = 0.5 g\Lambda\).

\begin{figure}
    \centering
         \includegraphics[width=0.45\textwidth]{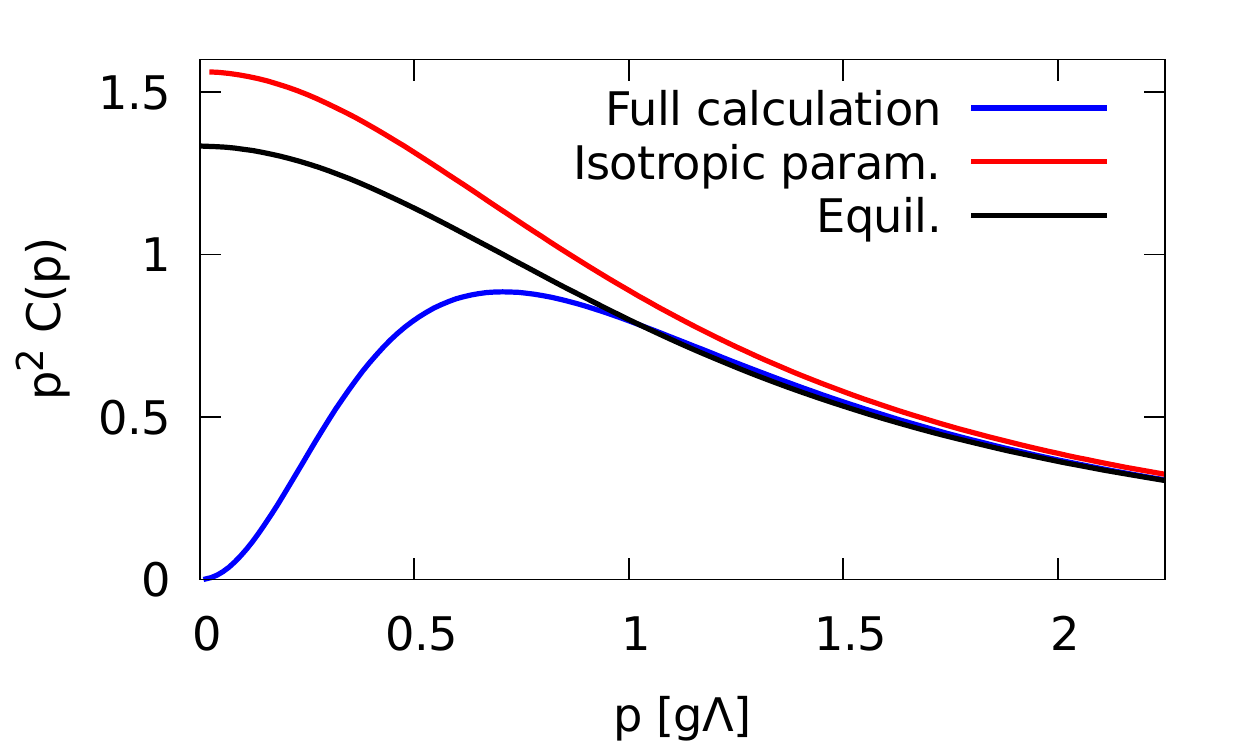}
         \caption{Comparison between the isotropic parametrization used in effective kinetic theory and our anisotropic collision kernel for \(\xi=-0.5\) and  \(\theta=\pi/2\).  The isotropic parametrization does not capture the anisotropic behaviour and incorrectly shows increased momentum broadening. The comparison for other values of \(\xi\) and \(\theta\) is similar.}
         \label{Fig:EKT}
\end{figure}

A non-equilibrium collision kernel is needed to calculate the rate of quasiparticle one-to-two radiation in kinetic theory \cite{Arnold:2002zm}. In the absence of an anisotropic non-equilibrium collision kernel, kinetic theory simulations have used an isotropic parametrization \cite{Kurkela:2015qoa, AbraaoYork:2014hbk} which is exact in an isotropic medium and  is given by Eq. \eqref{Eq:eq_coll_kern} where the equilibrium Debye mass is replaced by  
\beq
m_D^2 = 2 \int \frac{d^3 p}{(2\pi)^3 p}\;\left[2N_f f_q(\mathbf{p}) + 2N_c f_g(\mathbf{p}) \right]
\eeq
and the temperature is replaced by an effective temperature 
\beq
T_{*} =\frac{\frac{1}{2} \int \frac{d^3 p}{(2\pi)^3} \left[ 2N_f f_q(\mathbf{p}) (1-f_q(\mathbf{p})) + 2N_c f_g(\mathbf{p}) (1+f_g(\mathbf{p}))\right]}{\int \frac{d^3 p}{(2\pi)^3 p}\;\left[2N_f f_q(\mathbf{p}) + 2N_c f_g(\mathbf{p}) \right]} ,
\eeq
see \cite{Arnold:2002zm}.
In Fig.  \ref{Fig:EKT}, we compare this parametrization, as well as the equilibrium collision kernel at temperature \(T = \Lambda\), with our full anisotropic calculation of the collision kernel.  Clearly, the isotropic parametrization does not capture the behaviour of the collision kernel in an anisotropic medium. Including a full anisotropic collision kernel could affect the rate of quasiparticle splitting in kinetic theory simulations, and this influences the entire space-time evolution.

\subsection{Angular dependence of \(\hat{q}\)}
In an anisotropic medium, momentum broadening is distorted with more broadening in one direction of the transverse plane than the other. In other words, the collision kernel \(\mathcal{C}(\mathbf{q}_{\perp})\) depends on the direction of \(\mathbf{q}_{\perp}\). One way to quantify the total transverse momentum broadening in a particular direction  is to consider
\beq
\label{Eq:qhat_angular}
\begin{split}
\widehat{q}_{ij} &=  \int \frac{d^2 q_{\perp}}{(2\pi)^2}\; q_{\perp i} q_{\perp j}\,\mathcal{C}(\mathbf{q}_{\perp}).
\end{split}
\eeq
There always exist orthogonal principal axes so that \(\widehat{q}_{xy} = 0\). In our case, one axis is in the plane spanned by the jet direction and the anisotropy vector, with the other one being orthogonal. The total transverse momentum broadening in Eq. \eqref{Eq:qhat_def} is given by \(\widehat{q} = \widehat{q}_{xx} + \widehat{q}_{yy}\) and in general \(\widehat{q}_{xx} \neq \widehat{q}_{yy}\).

Eq. \eqref{Eq:qhat_angular} is UV divergent and needs a cutoff. This cutoff depends on the process being considered. As an example the cutoff for radiation off a highly energetic jet is roughly
\beq
q_{\mathrm{max}} \sim g\Lambda \sqrt{E/\Lambda}
\eeq
where \(E\) is the energy of the jet parton \cite{Arnold:2008zu} and \(\Lambda\) is the medium scale. This scaling relation becomes progressively better as the energy of the jet parton increases.

\begin{figure}
    \centering
    \begin{subfigure}[b]{0.45\textwidth}
         \centering
         \includegraphics[width=\textwidth]{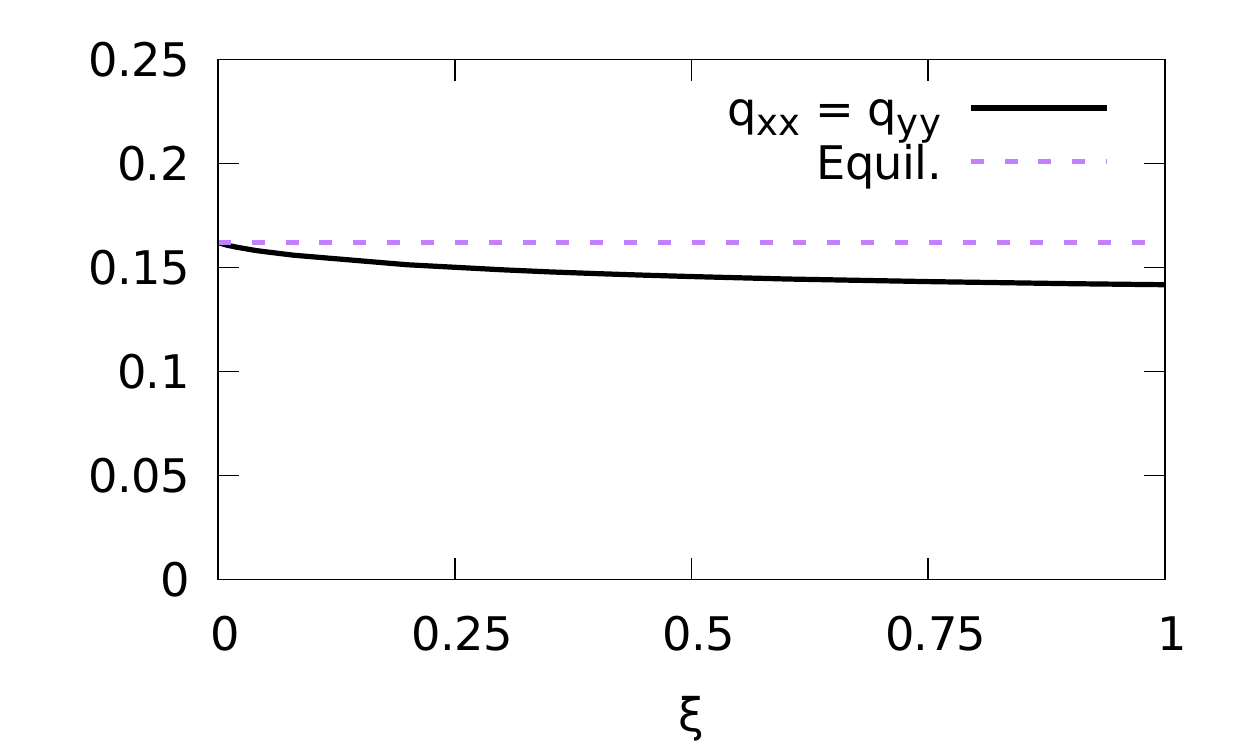}
         \caption{UV cutoff of \(q_{\mathrm{max}} = 3.2 g \Lambda\) corresponding roughly to jet energy \(E \sim 10 \Lambda\).}
         \label{Fig:}
     \end{subfigure}
     \hfill
    \begin{subfigure}[b]{0.45\textwidth}
         \centering
         \includegraphics[width=\textwidth]{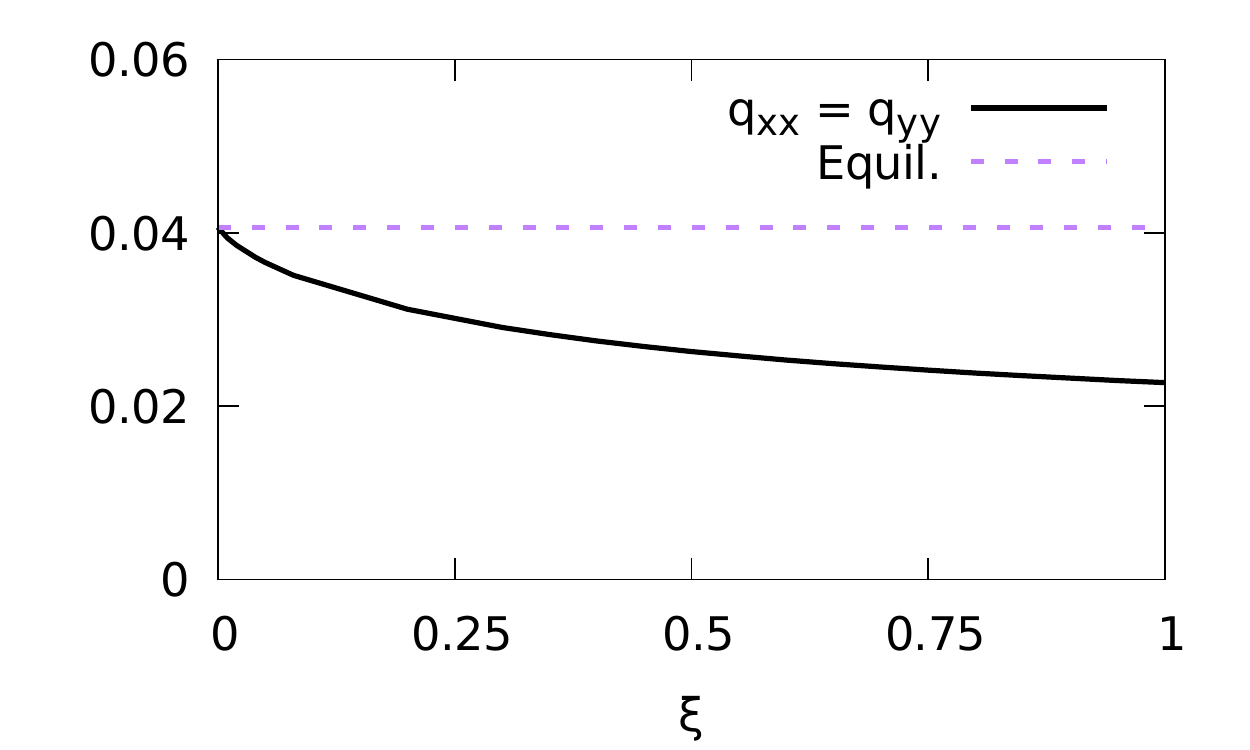}
         \caption{\(q_{\mathrm{max}} = g \Lambda\), corresponding roughly to \(E \sim \Lambda\) }
         \label{Fig:qmax_theta0_b}
     \end{subfigure} 
    \caption{The transport coefficients \(q_{xx}\) and \(q_{yy}\) in units of \(g^4 \Lambda^3\) in a medium with positive anisotropy. The jet parton travels parallel to the medium anisotropy, \(\theta = 0\). }
\label{Fig:qmax_theta0}
\end{figure}

In Fig. \ref{Fig:qmax_theta0}, we show transverse momentum broadening for a jet parton that travels in the direction of the medium anisotropy, i.e. \(\theta = 0\). In this case the transverse plane is the same in all directions and \(\widehat{q}_{xx} = \widehat{q}_{yy}\).  We consider a medium with positive anisotropy. As the energy of a jet partons increases it receives transverse kicks of higher energy which are less sensitive to details of medium screening. Therefore, anisotropy has more effect on low-energy jet partons. As an example, for a jet parton with \(E \sim 10 \Lambda\), the anisotropy gives a modest decrease of \(15 \%\) to momentum broadening, while for medium particles or very low-energy jet partons with \(E \sim \Lambda\), the decrease is nearly  \(45 \%\), see Fig. \ref{Fig:qmax_theta0}.

\begin{figure}
    \centering
    \begin{subfigure}[b]{0.45\textwidth}
         \centering
         \includegraphics[width=\textwidth]{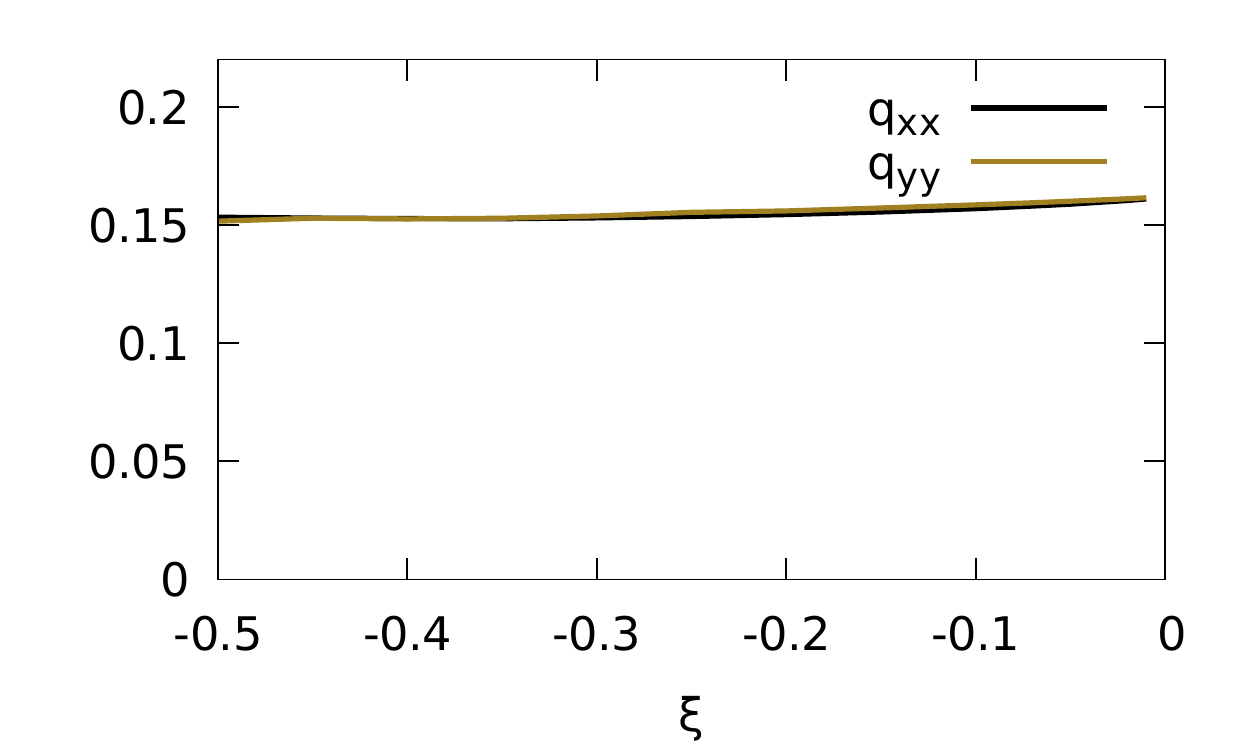}
         \caption{\(q_{\mathrm{max}} = 3.2 g \Lambda\) corresponding roughly to \(E \sim 10 \Lambda\).}
         \label{Fig:}
     \end{subfigure}
     \hfill
    \begin{subfigure}[b]{0.45\textwidth}
         \centering
         \includegraphics[width=\textwidth]{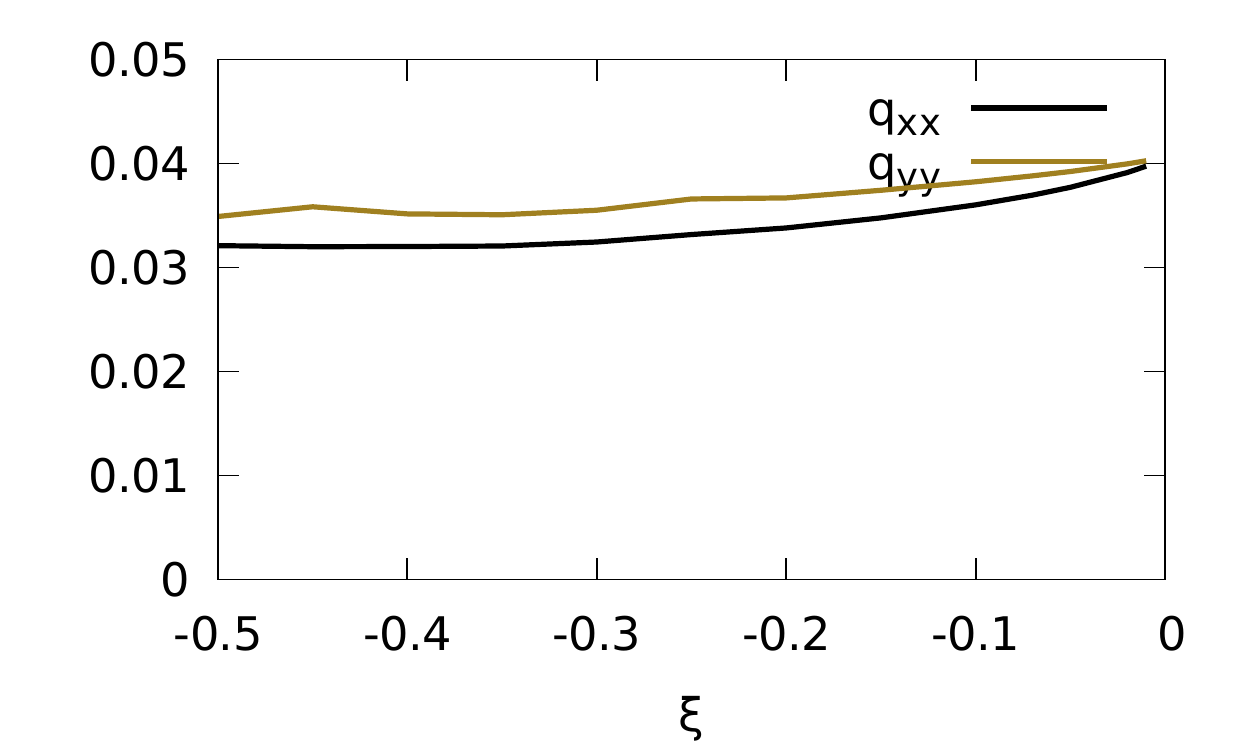}
         \caption{\(q_{\mathrm{max}} = g \Lambda\), corresponding roughly to \(E \sim \Lambda\) }
         \label{Fig:qmax_thetahalfpi_b}
     \end{subfigure} 
    \caption{The transport coefficients \(q_{xx}\) and \(q_{yy}\) in a medium with negative anisotropy. The jet parton is orthogonal to the medium anisotropy, \(\theta = \pi/2\). }
\label{Fig:qmax_thetahalfpi}
\end{figure}

In Fig. \ref{Fig:qmax_thetahalfpi}, we show transverse momentum broadening for a jet parton that travels orthogonally to the medium anisotropy, i.e. \(\theta = \pi/2\). We assume a medium with negative anisotropy. The effect of the anisotropy for jet parton with energy \(E \sim 10 \Lambda\) is small. However, for energy \(E \sim \Lambda\) there is a decrease of around \(20\%\) for \(q_{xx}\) and around \(15\%\) for \(q_{yy}\) due to the anisotropy. Thus there is modest distortion in transverse broadening anisotropic.

The effects of anisotropy on momentum broadening are most substantial for low-energy particles such as medium particles. Therefore, the effects we have described should be particularly important for quasiparticle interaction in kinetic theory as well for photon radiation from a medium. In these cases a medium quark with \(E\sim \Lambda\) receives transverse momentum kicks, causing it to radiate a gluon or a photon. Figs. \ref{Fig:qmax_theta0_b} and \ref{Fig:qmax_thetahalfpi_b} give a rough indicator of how important the effect of anisotropy on momentum broadening could be for the rate of those processes. In particular, Fig. \ref{Fig:qmax_theta0_b}  suggests that the effect could be sizable. In this regime one must solve the full equations for rate of gluon and photon radiation derived in \cite{Arnold:2001ba,Arnold:2002ja} which require the full collision kernel \(\mathcal{C}(\mathbf{q}_{\perp})\). We will explore this further in a future publication.


\section{Conclusions}

The rate of transverse momentum broadening is an integral part of the physics of relativistic plasmas. This rate is given by the collision kernel \(\mathcal{C}(\mathbf{q_{\perp}})\), the probability that a jet parton gets transverse kicks of momentum \(\mathbf{q}_{\perp}\) from the medium. The collision kernel also describes momentum broadening of a quasiparticle in the medium. 

For the first time, we have calculated the collision kernel in a non-equilibrium plasma, see Eq. \eqref{Eq:coll_kern}. This is motivated by the quark-gluon plasma in heavy-ion collsions which is anistropic and at times far from equilibrium. We assume an anisotropic momentum distribution of quasiparticles and calculate the \(rr\) correlator of soft gluons in the medium, see Eq. \eqref{Eq:rr_total_aniso}. Even at small values of anisotropy, the collision kernel changes qualitatively from the equilibrium result, see Figs. \ref{Fig:dep_on_xi_positive} and \ref{Fig:dep_on_xi_negative}. In those figures, additional screening leads to less momentum broadening, especially at small and intermediate transverse momenta, an effect which is not captured by the isotropic ansatz used in kinetic theory simulations \cite{Kurkela:2015qoa, AbraaoYork:2014hbk}. This gives a sizable decrease in the transport coefficient \(\widehat{q}\), as well as mild angular dependence, see Figs. \ref{Fig:qmax_theta0} and \ref{Fig:qmax_thetahalfpi}.

The collision kernel for momentum broadening determines the rate of medium-induced gluon radiation off a jet parton. It also determines the rate of one-to-two scattering of quasiparticles in a medium, as well as the rate of photon production through bremsstrahlung. With an anisotropic collision kernel in hand, we can therefore calculate consistently the rate of all these different processes in an anisotropic plasma. We will report on progress in a future publication.


A central challenge in this calculation is the presence of unstable modes in an anisotropic plasma. These modes lead to exponential growth in soft gluon density at very early times but numerical simulations suggest that at later times the modes become saturated  in heavy-ion collisions \cite{Berges:2013eia,Berges:2013fga,Berges:2014bba}. This is not captured by our analytic calculation,  leading us to subtract ultrasoft instability modes with exponential growth. 
Our focus is thus on fluctuating modes which are sourced at each instant by quasiparticles in the medium and whose contribution to momentum broadening does not depend on the detailed history of the medium. However, we note that for certain values of the anisotropy and jet direction, the collision kernel is sensitive to the ultrasoft modes subtracted and thus our calculation needs to be complemented by a description of the deep infrared in heavy-ion collisions. An alternative would be to measure the collision kernel defined in terms of Wilson lines \cite{Casalderrey-Solana:2007ahi} directly from, say, classical-statistical simulations of heavy-ion collisions.

\acknowledgements
This work was supported in part by the Natural Sciences and Engineering Research Council of Canada. S. H. gratefully acknowledges a scholarship from the Fonds de recherche du Qu\'ebec - Nature et technologies.

\bibliography{references}

%
%


\appendix

\section{Estimate of dependence on momentum cutoff}
\label{sec:Appendix_cutoff}

We will now show in detail the logarithmic divergence that occurs due to instability modes when momentum broadening is evaluated as in Sec. \ref{sec:Density of soft gluons} making the (incorrect) assumption of an initial condition at time \(t_0 = -\infty\).
 For simplicity, we will consider the case where the jet parton momentum \( \widehat{\mathbf{k}}\) is parallel to the anisotropy vector \(\mathbf{n}\) which we choose to be the \(z\) axis. Momentum broadening is given by 
\beq
\label{Eq:qhat_app}
\widehat{q} \approx  g^2 C_F \int \frac{d^4 Q}{(2\pi)^4}\; q^2_{\perp} D_{rr}^{\mu\nu}(Q) v_{\mu} v_{\nu} \,\delta(v\cdot Q).
\eeq
where 
\beq
D_{rr} = D_{\ret} \Pi_{aa} D_{\adv}.
\eeq
in this naive setup.
The blowup due to instabilities comes from the instability poles in \(D_{\ret}\) and \(D_{\adv}\). We focus on the instability poles coming from the term with \(\Pi_e\). The other contribution can be evaluated similarly. Using results from \cite{Romatschke:2003vc,Carrington:2014bla} one can show that at small anisotropy 
\beq
\begin{split}
\label{Eq:Pie_app}
\Pi_{e} &= m_D^2 \left(\frac{\omega}{q}\right)^2 - \frac{i\pi}{4} m_D^2 \frac{\omega}{q} + \mathcal{O}\left(\left(\frac{\omega}{q}\right)^3\right) \\
& +\xi \left[-\frac{1}{6} (1+\cos 2\theta)m_D^2 + \mathcal{O}\left( \frac{\omega}{q}\right) \right].
\end{split}
\eeq
We verify below that the expansion in \(\omega/q \ll 1\) is consistent.
This clearly shows the presence of instability poles as
\beq
\omega^2 = q^2 + \Pi_e \approx q^2 - \xi \frac{1}{6} (1+\cos 2\theta)m_D^2
\eeq
has an imaginary solution \(\omega = i\gamma\) for sufficiently small values of \(q\). For our simple estimate we can ignore the dependence of the pole on the angle \(\theta\) between the gluon momentum \(\mathbf{q}\) and the anisotropy vector \(\mathbf{n}\).

Ignoring all numerical factors and substituting \(D_{\ret} \rightarrow 1/(Q^2-\Pi_e)\) and \(D_{adv} = D_{\ret}^{*}\) in Eq. \eqref{Eq:qhat_app}, we get
\beq
\begin{split}
\widehat{q} &\sim g^3 \Lambda^2 \int d^4 Q q_{\perp}^2 \delta(\omega - q^z) \\
&\times \frac{1}{(q_{\perp}^2 + m_D^2\left(\frac{\omega}{q}\right)^2 - \xi m_D^2)^2 + \left( m_D^2 \frac{\omega}{q}\right)^2}
\end{split}
\eeq
It is clear that this expression diverges when \(q_{\perp} \sim \sqrt{\xi} m_D\) and \(\omega \sim q_{\perp}^3/m_D^2 \sim \xi^{3/2} m_D\), justifying our approximation that \(\omega/q \ll 1\) in Eq. \eqref{Eq:Pie_app} when the anisotropy is small. Rewriting gives
\beq
\widehat{q} \sim g^3 \Lambda^2 \int dq^z \int dq_{\perp} \frac{q_{\perp}^3}{(\frac{m_Dq^z}{\sqrt{\xi}})^2 + \left( q_{\perp}^2 - \xi m_D^2\right)^2}
\eeq
Scaling all variables by \(m_D\) and introducing a cutoff \(\delta\) defined by \(q_{\perp} \approx q > \sqrt{\xi} m_D + \delta m_D\) gives that
\beq
\widehat{q} \sim g^4 \Lambda^3 \int dx \int_{\sqrt{\xi} + \delta } dz\; \frac{z^3}{x^2/\xi + (z^2 -\xi)^2}
\eeq
The remaining integrals can be done analytically by a change of variables \(x \rightarrow x/\sqrt{\xi}\) and \(y = z^2 - \xi\), and then going to radial coordinates. The result is that dependence of momentum broadening on the cutoff is
\beq
\label{Eq:dep_on_delta}
\widehat{q} \sim g^4 \Lambda ^3 \xi^{3/2} \log \left( \sqrt{\xi}\delta\right).
\eeq

Our derivation of the logarithmic blowup in Eq. \eqref{Eq:dep_on_delta} was done in a naive framework where the initial condition is at \(t_0 = -\infty\). However, this derivation also tells us about the case where the initial condition is specified at \(t_0 = 0\), see Sec. \ref{sec:instabilities}. When specifying a cut \(\omega_{\mathrm{cut}}\), below which instability poles are subtracted, the treatment for modes of energy greater than \(\omega_{\mathrm{cut}}\) is the same as in the derivation of  Eq. \eqref{Eq:dep_on_delta}. Given a relation between the three-momentum cutoff \(\delta\) and the frequency cutoff \(\omega_{\mathrm{cut}}\), we therefore see that the dependence on the frequency cutoff is
\beq
\widehat{q} \sim g^4 \Lambda ^3 \xi^{3/2} \log \left( \sqrt{\xi}\,\omega_{\mathrm{cutoff}}\right).
\eeq

\section{Gluon self-energy in an anisotropic medium}
\label{sec:Appendix_selfenergy}

To evaluate the \(rr\) correlator we need the self-energy \(\Pi_{aa}\) given by Eq. \eqref{Eq:Pi_aa}.  Doing a change of variables to \(\tilde{p} = p\sqrt{1+\xi \left(\mathbf{v}\cdot \mathbf{n}\right)^2}\) shows that
\beq
\Pi^{\mu\nu}_{aa} = 2\pi m_{D,\mathrm{eq}}^2\, \Lambda P^{\mu\nu} 
\eeq
where the angular part is 
\beq
\label{Eq:P_angular}
P^{\mu\nu} = \left. \int\frac{d\Omega}{4\pi}\; v^\mu v^\nu \frac{\delta(\omega - \mathbf{v}\cdot \mathbf{q})}{\left( 1+\xi  \left(\mathbf{v}\cdot \mathbf{n}\right)^2\right)^{3/2}} \, \right\vert_{v^0 = v}.
\eeq
Furthermore,
\beq
\begin{split}
 m_{D,\mathrm{eq}}^2 =  &\frac{g^2}{2\pi^2} \int_0^{\infty} d\tilde{p}\;\tilde{p}^2 \\ &\times \left[2N_f f^0_q (1-f^0_q) + 2N_c f^0_g (1+f^0_g) \right]
 \end{split}
\eeq
is the Debye mass in thermal equilibrium with \(f^0_q\) and \(f^0_g\) the equilibrium distributions with the temperature substituted by \(\Lambda\). Using the delta function in Eq. \eqref{Eq:P_angular} we get that
\beq
P^{00} = \frac{1}{4\pi q} 
\int_0^{2\pi}d\phi\;
\frac{1}{\left[ 1+\xi \left( \tilde{q}_{\parallel} \tilde{\omega} - \tilde{q}_{\perp} \sqrt{1-\tilde{\omega}^2} \cos \phi\right)^2\right]^{3/2}},
\eeq
\beq
P^{0i}n^i = \frac{1}{4\pi q} \int_0^{2\pi}d\phi\; \frac{ \tilde{q}_{\parallel} \tilde{\omega} - \tilde{q}_{\perp} \sqrt{1-\tilde{\omega}^2} \cos \phi}{\left[ 1+\xi \left( \tilde{q}_{\parallel} \tilde{\omega} - \tilde{q}_{\perp} \sqrt{1-\tilde{\omega}^2} \cos \phi\right)^2\right]^{3/2}},
\eeq
\beq
P^{ij}n^i n^j = \frac{1}{4\pi q} \int_0^{2\pi}d\phi\; \frac{ \left(\tilde{q}_{\parallel} \tilde{\omega} - \tilde{q}_{\perp} \sqrt{1-\tilde{\omega}^2} \cos \phi\right)^2}{\left[ 1+\xi \left( \tilde{q}_{\parallel} \tilde{\omega} - \tilde{q}_{\perp} \sqrt{1-\tilde{\omega}^2} \cos \phi\right)^2\right]^{3/2}},
\eeq
where \(\tilde{q}_{\parallel} = \mathbf{n}\cdot\mathbf{q}/q\) is the normalized component of the gluon momentum that is parallel to the anisotropy vector \(\mathbf{n}\), \(\tilde{q}_{\perp} = \left| \mathbf{q} - \left( \mathbf{n}\cdot\mathbf{q}\right)\mathbf{n}\right|/q \) is the normalized transverse component and \(\tilde{\omega} = \omega/q\). We do the remaining integral numerically. See \cite{Kasmaei:2018yrr} for an alternative evaluation, partially in terms of special functions.

We can finally assemble the components of \(\Pi_{aa}\) found in Eq. \eqref{Eq:Piaa_decomp},
\beq
-i \Pi_{aa} = \alpha P_L + \beta E + \gamma C + \delta D
\eeq
They are
\begin{align}
i \alpha &= 2\pi\Lambda\, m_{D,\mathrm{eq}}^2 \,\frac{\omega^2-q^2}{q^2}\,P^{00}, \\
i \delta &=   2\pi\Lambda\, m_{D,\mathrm{eq}}^2\,\frac{\omega/q}{\sqrt{q^2-\left( \mathbf{q} \cdot \mathbf{n}\right)}} \nonumber \\
&\times\left(P^{0i}n^i - \frac{\mathbf{q}\cdot\mathbf{n}\, \omega}{q^2} P^{00} \right), \\
i \gamma &=   2\pi\Lambda\, m_{D,\mathrm{eq}}^2\,\frac{1}{1-\left( \mathbf{q} \cdot \mathbf{n}\right)^2/q^2} \nonumber \\
&\times\left(P^{ij}n^i n^j - \frac{2 \mathbf{q}\cdot\mathbf{n}\, \omega}{q^2} P^{0i}n^i + \frac{(\mathbf{q}\cdot\mathbf{n})^2\omega^2}{q^4} P^{00} \right) \\
i\beta &= -i\alpha-i\gamma.
\end{align}

For the convenience of the reader we also collect the components of the retarded self-energy in Eq. \eqref{Eq:Piret_decomp},
\beq
-i \Pi^{\mu\nu}_{\ret} = \Pi_L P_L^{\mu\nu} + \Pi_e E^{\mu\nu} + \Pi_c C^{\mu\nu} + \Pi_d D^{\mu\nu}
\eeq
They were first derived in \cite{Romatschke:2003ms}. They have the same form as the components of \(\Pi_{aa}\) except that \(2\pi\Lambda m_{D,\mathrm{eq}}^2\) is replaced by \(\frac{1}{2} m_{D,\mathrm{eq}}^2\) and the angular integrals become
\begin{align}
P_{\ret}^{00} &= \int_{-1}^{1} dz \; \frac{1}{\left( 1+\xi z^2\right)^2} \nonumber
\\
& \times \Big[ -1 +\left(\omega + \xi q_{\parallel} z \right) R(\omega - q_{\parallel} z, q_{\perp}\sqrt{1-z^2})\Big], \\
P_{\ret}^{0i} n^i &= \int_{-1}^{1} dz \; \frac{z}{\left( 1+\xi z^2\right)^2}  \nonumber \\
&\times\Big[ -1 +\left(\omega + \xi q_{\parallel} z \right) R(\omega - q_{\parallel} z, q_{\perp}\sqrt{1-z^2})\Big], \\
P_{\ret}^{ij}n^i n^j &= \frac{1+\xi}{\xi^{3/2}}\left( \arctan \sqrt{\xi} - \frac{\sqrt{\xi}}{1+\xi}\right) \nonumber \\
&+\int_{-1}^{1} dz \; \frac{z^2}{\left( 1+\xi z^2\right)^2} \nonumber \\
&\times \Big[ -1 + \left(\omega + \xi q_{\parallel} z \right) R(\omega - q_{\parallel} z, q_{\perp}\sqrt{1-z^2})\Big]
\end{align}
where
\beq
\begin{split}
R(a,b) &= \int_0^{2\pi} \frac{d\phi}{2\pi} \frac{1}{a-b\cos \phi + i \epsilon} \\
&= \theta(a^2-b^2) \frac{\mathrm{sgn}(a)}{\sqrt{a^2-b^2}} - \theta(b^2-a^2) \frac{i}{\sqrt{b^2-a^2}}
\end{split}
\eeq
Furthermore, 
\beq
\Pi_e = - \Pi_L - \Pi_c + \frac{\arctan \sqrt{\xi}}{\sqrt{\xi}}m_{D,\mathrm{eq}}^2
\eeq



\section{Energy loss in an unstable medium}
\label{sec:Appendix_energyloss}

In Eq. \eqref{Eq:ret_final} we showed that a contour \(\alpha\) that goes above all instability poles in the frequency domain is needed to transform the retarded correlator to the time domain \cite{Hauksson:2020wsm}. Assuming the factorization in Eq. \eqref{Eq:ret_scale_sep} this gives that
\beq
\begin{split}
&G_{\ret}(t_x-t_y, \mathbf{k}) =  \\
&\int \frac{dp^0}{2\pi} \;e^{-ip^0(t_x-t_y)} \widehat{G}_{\ret}(p^0,\mathbf{p}) + \theta(t_x - t_y) \sum_{i} A_i e^{\gamma_i (t_x-t_y)}
\end{split}
\eeq
which guarantees that \(G_{\ret}(t_x,t_y) = 0\) for \(t_x<t_y\), i.e. that all propagation is causal and that instabilities give exponential growth. 
If we were to use an incorrect contour along the real line, the retarded correlator in the time domain would become
\beq
\label{Eq:ret_wrong}
\int \frac{dp^0}{2\pi} \;e^{-ip^0(t_x-t_y)} \widehat{G}_{\ret}(p^0,\mathbf{p}) + \theta(t_y - t_x) \sum_{i} A_i e^{\gamma_i (t_x-t_y)}.
\eeq
This is non-vanishing for \(t_x < t_y\) and thus does not describe causal evolution.

Earlier work on heavy-quark energy loss in an anisotropic plasma \cite{Romatschke:2003vc, Romatschke:2004au} implicitly used a contour along the real line, i.e. Eq. \eqref{Eq:ret_wrong}. An easy way to see this is to note that their end result for energy loss was Eq. \eqref{Eq:e_loss} which we derived using regular Fourier transforms and not the contour \(\alpha\). Alternatively, the origin of the incorrect contour can be understood using the classical derivation in Eq. \eqref{Eq:e_loss_class_der}.  This relied on Eq. \eqref{Eq:linear_response} from linear response theory which describes how an electric field is induced by a current. The linear response relation comes from Fourier transforming the fundamental relation for the electric field,
\beq
\label{Eq:inst_linear_resp}
\begin{split}
&E^i(X) = \partial_{x^0} \int d^4 Y \; D^{ij}_{\ret}(X-Y) J^j_{\mathrm{ext}}(Y) \\
&=\partial_{x^0} \int d^4 Y \; \int \frac{d^4 Q}{(2\pi)^4} \,e^{-i(X-Y)\cdot Q} D^{ij}_{\ret}(Q) \int \frac{d^4 K}{(2\pi)^4} \,e^{-iK \cdot Y} J^j_{\mathrm{ext}}(K).
\end{split}
\eeq
Doing the \(Y\) integral then gives a delta function \(\delta^{(4)}(Q-K)\), leading to Eq. \eqref{Eq:linear_response}. However, in a system with an instability, things are not so simple. To preserve causality, we must use the contour \(\alpha\) for the \(q^0\) integral in Eq. \eqref{Eq:inst_linear_resp}. Thus \(q^0\) can be in the upper half complex plane. For an instability pole \(q^0 = i\gamma\), the integral is
\beq
\label{Eq:app_divergence}
\int^{x^0}_{-\infty} dy^0 e^{-i(x^0-y^0)q^0} = \int_{-\infty}^{x^0} dy^0 e^{(x^0-y^0)\gamma} = \infty.
\eeq
which is divergent. 
The interpretation here is straightforward. Assuming a plasma that has existed for a long time compared to the rate of energy loss, the jet parton will have sourced unstable fields a long time ago that will have had time to grow to very large amplitudes leading to extremely large energy loss. Thus using the setup of \cite{Romatschke:2003vc, Romatschke:2004au} should give infinite energy loss due to instabilities. It should be emphasized that the calculations of heavy quark energy loss in \cite{Romatschke:2003vc, Romatschke:2004au} correctly got the involved contribution of fluctuating soft modes and hard scattering.

The correct procedure in an unstable plasma is to use equations in the time domain like in the first line of Eq. \eqref{Eq:inst_linear_resp} and start the system at some finite time \(t_0 = 0\). Then the integral in Eq. \eqref{Eq:app_divergence} is 
\beq
\int_{0}^{x^0} dy^0 e^{(x^0-y^0)\gamma} = \frac{e^{\gamma x^0} -1}{\gamma}.
\eeq
For later times \(t_x\) the system will have evolved and the retarded correlator should be modified. This requires numerical calculations and has been performed in e.g. \cite{Carrington:2015xca,Mrowczynski:2017kso}. Alternatively, if one is simply interested in fluctuating modes such as in this work, one could subtract instability poles below a cutoff \(\omega_{\mathrm{cut}}\).

Other earlier calculations of probes in an anisotropic plasma do not suffer from this flaw. Calculation of photon emission through two-to-two scattering, see e.g. \cite{Schenke:2006yp}, only use a resummed quark propagator which has no instability poles \cite{Schenke:2006fz}. Furthermore, the heavy-quark potential \cite{Dumitru:2007hy} describes equal-time correlators and thus there is no room for instabilities to grow.

\section{Correction to \cite{Hauksson:2020wsm}}
\label{sec:App_correction}

In our earlier paper \cite{Hauksson:2020wsm}, we showed that the \(rr\) propagator in an unstable plasma is 
\begin{widetext}
\beq
\label{Eq:rr_result_app}
\begin{split}
G_{rr}(x^0,y^0) \approx &\int \frac{dk^0}{2\pi}\; \widehat{G}_{\ret}(k^0) \,\Pi_{aa}(k^0)\, \widehat{G}_{\adv}(k^0) \;\;e^{-ik^0(x^0-y^0)} \\
+ \sum_i &\int \frac{dk^0}{2\pi}\; \frac{A_i}{k^0-i \gamma_i} \,\Pi_{aa}(k^0)\, \widehat{G}_{\adv}(k^0) \;\left(e^{-ik^0x^0}-e^{\gamma_i x^0} \right) e^{ik^0y^0} \\
+ \sum_j &\int \frac{dk^0}{2\pi}\; \widehat{G}_{\ret}(k^0) \,\Pi_{aa}(k^0)\, \frac{A_j^*}{k^0+i \gamma_j} \;e^{-ik^0x^0} \left( e^{ik^0y^0}-e^{\gamma_j y^0}\right) \\
+ \sum_{i,j} &\int \frac{dk^0}{2\pi}\; \frac{A_i}{k^0-i \gamma_i} \,\Pi_{aa}(k^0)\, \frac{A_j^*}{k^0+i\gamma_j}  \; 
\left(e^{-ik^0x^0}-e^{\gamma_i x^0} \right)\left( e^{ik^0y^0}-e^{\gamma_j y^0}\right),
\end{split}
\eeq
\end{widetext}
assuming the factorization in Eq. \eqref{Eq:ret_scale_sep}. The first term describes fluctuating modes, the last term describes instability modes and the middle terms are cross-terms between fluctuating and instability modes. Using a set of controlled approximations, Eq. \eqref{Eq:rr_result_app} can be shown to be equivalent to Eq. \eqref{Eq:rr_final} \cite{Hauksson:2020wsm}.

In \cite{Hauksson:2020wsm}, we gave a heuristic discussion of Eq. \ref{Eq:rr_result_app}. This discussion, given in Eqs. (37) to (39) of  \cite{Hauksson:2020wsm}, contained some wrong signs which we correct here. We emphasize that all results of  \cite{Hauksson:2020wsm}, including Eq. \eqref{Eq:rr_result_app}, remain unaffected.

The \(rr\) propagator can be written as 
\begin{widetext}
\beq
G_{rr}(x^0,y^0,\mathbf{k}) 
= \int dw^0 \int dz^0\; G_{\ret}(x^0,w^0;\mathbf{k}) \Pi_{aa}(w^0,z^0;\mathbf{k}) G_{\adv}(z^0,y^0;\mathbf{k}).
\eeq
\end{widetext}
Omitting dependence on the three-momentum \(\mathbf{k}\), this is
\beq
\label{Eq:rr_app_full}
\begin{split}
G_{rr}(x^0,y^0) = \int dw^0 &\int dz^0\int \frac{dk^0}{2\pi} \; G_{\ret}(x^0,w^0)   \\
&\times\, e^{-ik^0(w^0-z^0)}\,\Pi_{aa}(k^0) G_{\adv}(z^0,y^0)
\end{split}
\eeq
where we Fourier transformed the self-energy \(\Pi_{aa}\). This equation can easily be evaluated schematically. A mode in the retarded propagator with energy \(E\) and decay rate \(\Gamma\) is  
\beq
\label{Eq:mode_ret}
G_{\ret}(x^0,w^0) \sim \theta(x^0-w^0) e^{-iE(x^0-w^0)-\Gamma (x^0-w^0)}
\eeq
in the time domain. In thermal equilibrium, or generally any system starting at time \(t_0 = -\infty\), this contributes
\beq
\int^{x^0}_{-\infty} dw^0 \;e^{-ik^0w^0} e^{-i(E-i\Gamma)(x^0-w^0)} = \frac{i e^{-ik^0 x^0}}{k^0-E+i\Gamma}
\eeq
to the \(rr\) correlator in Eq. \eqref{Eq:rr_app_full}. However, in a system that is started at time \(t_0 = 0\), the contribution of this mode is
\beq
\begin{split}
&\int^{x^0}_{0} dw^0 \;e^{-ik^0w^0} e^{-i(E-i\Gamma)(x^0-w^0)}\\
&= \frac{i}{k^0-E+i\Gamma} \left[  e^{-ik^0 x^0} - e^{-i(E-i\Gamma)x^0}\right].
\end{split}
\eeq

Generally speaking, for a pole \(b\) in the retarded propagator, we should have a contribution
\beq
\frac{1}{k^0-b}\left( e^{-ik^0x^0} - e^{-ib x^0}\right)
\eeq 
in Eq. \eqref{Eq:rr_result_app}. 
However, controlled approximation allow us to drop terms \(e^{-ibx^0}\) when \(b\sim g\Lambda\) since those terms either decay rapidly or oscillate too fast to contribute to momentum broadening \cite{Hauksson:2020wsm}. This explains the form of \eqref{Eq:rr_result_app}: For instability poles \(b = i\gamma\), we get a contribution
\beq
\frac{1}{k^0-i\gamma} \left( e^{-ik^0x^0} - e^{\gamma x^0}\right)
\eeq
while for poles with \(b\sim g\Lambda\), we simply get a contribution 
\beq
\frac{1}{k^0-b} e^{-ik^0x^0}.
\eeq

\end{document}